\documentclass[reprint,amsmath,amssymb,aps,nofootinbib]{revtex4-2}
\usepackage{graphicx}
\usepackage{dcolumn}
\usepackage{bm}
\usepackage{physics}
\usepackage{hyperref}
\hypersetup{
     colorlinks=true,
     linkcolor=blue,
     filecolor=blue,
     citecolor = blue,      
     urlcolor=blue,
     }
\usepackage[super]{nth}
\usepackage{physics}

\usepackage[shortlabels]{enumitem}
\usepackage{bbm}
\usepackage{orcidlink}

\newcommand{\gi}{\ensuremath{g^{(i)}_{\mu\nu}}}
\newcommand{\Wi}{\ensuremath{W^{(i)\mu}_{\;\;\;\;\;\;\nu}}}
\newcommand{\Gi}{\ensuremath{G^{(i)\mu}_{\;\;\;\;\;\;\nu}}}

\begin{document}

\title{Black Holes in Multi-Metric Gravity II: \\ Hairy Solutions and Linear Stability of the \\ Non- and Partially Proportional Branches}
\author{Kieran Wood, \orcidlink{0000-0002-4680-5563}}
\email{kieran.wood@nottingham.ac.uk}
\affiliation{School of Physics and Astronomy, University Of Nottingham, Nottingham NG7 2RD, UK}
\affiliation{Nottingham Centre Of Gravity, University Of Nottingham, Nottingham NG7 2RD, UK}

\author{Paul M. Saffin, \orcidlink{0000-0002-4290-3377}}
\email{paul.saffin@nottingham.ac.uk}
\affiliation{School of Physics and Astronomy, University Of Nottingham, Nottingham NG7 2RD, UK}
\affiliation{Nottingham Centre Of Gravity, University Of Nottingham, Nottingham NG7 2RD, UK}

\author{Anastasios Avgoustidis, \orcidlink{0000-0001-7247-5652}}
\email{anastasios.avgoustidis@nottingham.ac.uk}
\affiliation{School of Physics and Astronomy, University Of Nottingham, Nottingham NG7 2RD, UK}
\affiliation{Nottingham Centre Of Gravity, University Of Nottingham, Nottingham NG7 2RD, UK}

\begin{abstract}
    Owing to our work in part I of this series of papers, it is understood that the analytically known black hole solutions in the theory of ghost free multi-metric gravity can be split into three distinct classes, and that one of these classes -- the proportional branch -- exhibits the Gregory-Laflamme instability at linear level in the metric perturbations, whenever the black hole horizon size is smaller than (roughly) the Compton wavelength of the theory's lightest massive graviton. In this first of two sequels, we determine the linear stability of the two remaining classes of black hole solutions -- the non-proportional and partially proportional branches -- and discuss how our results likely differ at nonlinear level. We also give a general prescription to construct multi-metric solutions describing black holes endowed with massive graviton hair, which may constitute the end state of the instability in the proportional branch. We utilise a tractable example model involving 3 metrics to see how this works in practice, and determine the asymptotic form of its corresponding hairy solutions at infinity, where one can clearly see the individual contributions from each of the graviton mass modes.
\end{abstract}

\maketitle

\section{Introduction}

Theories of massive and multi-metric gravity have garnered considerable attention in recent years, owing to their ability to potentially say something meaningful about a number of problems at the interface between gravity and particle physics \cite{BHs_multigrav,dR_review,Hinterbichler_review,bigravity_review,interacting_spin2}. These theories extend general relativity (GR) by introducing additional interacting massive spin-2 fields over and above the single massless graviton of GR, whose presence can influence the gravitational dynamics in interesting and useful ways; for example, one can talk about possibly solving the hierarchy problem \cite{ClockworkGrav,Deconstructing,ClockworkCosmo}, and the theory may contain a viable dark matter candidate \cite{heavy_spin2_DM,bigravity_DM,DM_multigrav,Oscillating_DM}, amongst other niceties. Nonlinearly, the inclusion of these additional spin-2 fields manifests as a framework in which multiple metric tensors interact with one another on the same spacetime manifold (hence the name `multi-metric gravity', or sometimes just `multi-gravity'). 

As with any theory of modified gravity, these multi-metric theories must pass both observational and theoretical tests to at least as high a degree of precision as GR in order to be considered viable alternatives. It is, therefore, of paramount importance that we determine whether these theories are, at the very least, theoretically consistent, before we can then go on to investigate where they might give predictions that are testably different from GR. One such theoretical requirement is of course that a given theory should possess stable solutions that are able to describe real physical systems existing in our universe; the natural arena to probe in gravitational theories is that of black holes, since they do exist and provide a convenient window into the strong curvature regime in any classical theory of gravity. 

In part I of this series of papers \cite{BHs_multigrav} -- which we will recap in more detail in section \ref{Sec:recap} of this work -- we explicitly constructed a wide class of black hole solutions to the general theory of multi-metric gravity (in arbitrary spacetime dimension) and found that they can be separated into three distinct classes depending on whether the various metrics can be simultaneously diagonalised. We also showed that the simplest of these three solution branches (the proportional branch) can exhibit an instability whose defining equations are equivalent to those governing the Gregory-Laflamme instability afflicting black strings in theories with extra dimensions \cite{GL_instability,Charged_GL_instability,AdS_GL_instability}, depending on the mass of the theory's lightest graviton. Our results directly extended and generalised the analogous results already known in 4-dimensional dRGT massive gravity and bigravity \cite{consistent_spin2,prop_bg_multigrav,KoyamaNiz,Exact_schdS,dR_BHs,horizon_struct_bigravity,spherical_sym_sols,charged_BHs_bigravity,rotating_BHs_bigravity,Rotating_AdS_bigravity,GL_instability_bigravity,Kerr_instability_bigravity}, and shed some additional light on the nature of the instability and its relation to dimensional deconstruction \cite{discrete_grav_dims,deconstructing_dims,Deconstructing,ClockworkCosmo} -- we will have a little more to say on this in section \ref{sec:prop instability} as well.

This time, in part II, we determine the linear stability of the remaining two branches of analytically known black hole solutions (the non-proportional and partially proportional branches) and conjecture upon how our conclusions from the linear analysis may change dramatically if one were to consider nonlinear effects. Building on from part I, we also give a prescription to construct more generic spherically symmetric black hole solutions in these multi-metric theories that are posited to exist beyond the instability threshold in the proportional branch. That such solutions should exist can be understood purely on the grounds that once the instability ensues, it must necessarily saturate \emph{somewhere}, and if the initial state is spherically symmetric then so too should the final state be (unless one wishes for cosmic censorship to be violated). In practice, constructing these solutions is extremely difficult even for the simplest of toy models, and must be done numerically. Nevertheless, one can still understand the generic procedure one must follow in order to construct these solutions. Moreover, perhaps more usefully, one can glean physical information by considering the asymptotic behaviour of these solutions at infinity, where it can be clearly seen (analytically) that they describe black holes supplemented by massive graviton hair. Such hairy black hole solutions have been explicitly constructed already in bigravity \cite{Hairy_BH_AdS,Hairy_BHs_flat,Hairy_BHs_Gervalle}, where the solutions are built by analytically determining their asymptotic form at infinity and then numerically solving the boundary value problem defined by these asymptotics. The presence of additional metrics complicates matters significantly; indeed, we feel the full numerical calculation warrants a separate article that will form the final installment in this series of papers, but we will still provide a blueprint here that one may follow if one wishes to construct these solutions for a generic multi-metric model, as well as an explicit calculation of the asymptotics for an example model with three metrics that still contains the relevant physics.

The structure of this work, then, is as follows: in section \ref{Sec:recap} we review multi-metric theory and provide a recap of what we did in part I; in section \ref{Sec:stability} we determine the linear stability of the non-proportional and partially proportional branches of black hole solutions; in section \ref{Sec:hairy} we provide a general procedure by which one would construct the hairy black hole solutions for any multi-metric theory, and use it to determine the asymptotic form of one such solution in an example model with 3 metrics (we will save the numerics for part III); finally we conclude in section \ref{Sec:conclusion}. 

We work in natural units $c=\hbar=G=1$ throughout, and always use a mostly-plus metric signature.
 
\section{Recap of Part I: multi-metric theory, black hole solutions and instability of proportional branch}\label{Sec:recap}

We begin this first sequel with a synopsis of our precursor work \cite{BHs_multigrav}. The theory of multi-metric gravity can be formulated in two different but nevertheless equivalent ways: one may work in either the \emph{metric formalism}, where the potential governing the interactions between the various metrics is constructed from the metrics directly, or the \emph{vielbein formalism} (also known as the \emph{tetrad formalism}), where it is instead written in terms of a wedge product between the different tetrad 1-forms associated to each of the metrics. Both approaches have their benefits and drawbacks and are useful in different situations; in part I, we used them both to tackle different problems, though we still explained how one may convert between formalisms if one so wishes. In part II, we shall work exclusively in the metric formalism, since to discuss stability we will be concerned with linear perturbations around the background black hole solutions; in \cite{BHs_multigrav} we saw that the metric approach is most appropriate for this problem. Naturally, our intention is that parts I, II and eventually III be read in conjunction with one another, so if the reader wishes to know how the calculations we shall present here would look in the language of vielbeins, we direct them to \cite{BHs_multigrav}.

The starting point for all our analysis, then, is the metric version of the standard ghost free multi-metric action, $I$, in vacuum\footnote{We shall in the later sections restrict ourselves to work in $D=4$ dimensions, for simplicity of the stability analysis, but for the moment we shall maintain generality.}:
\begin{align}\label{MultigravAction}
    I &= I_K + I_V
    \\
    I_K &= \sum_{i=0}^{N-1} \frac{M_i^{D-2}}{2} \int \dd[D]x\, \sqrt{-\det g_{(i)}} R_{(i)}\label{MultigravKinetic}
    \\
    I_V &=  -\sum_{i,j}\int \dd[D]x\, \sqrt{-\det g_{(i)}} \sum_{m=0}^{D} \beta_m^{(i,j)} e_m(S_{i\rightarrow j}) \label{MultigravPotential} \; ,
\end{align}
where the kinetic term is simply $N$ copies of the Einstein-Hilbert action (one for each metric), and the interaction potential is built by summing the elementary symmetric polynomials $e_m$ of the eigenvalues of the characteristic building block matrices:
\begin{equation}\label{Sij}
    S_{i\rightarrow j} = \sqrt{g_{(i)}^{-1}g_{(j)}} \; ,
\end{equation}
along with some arbitrary constant parameters $\beta_m^{(i,j)}=\beta_m^{(j,i)}$ (of mass dimension $D$) to characterise the interactions between $g_{\mu\nu}^{(i)}$ and $g_{\mu\nu}^{(j)}$. In Eq. \eqref{Sij}, the matrix square root is defined in the sense that $(S^2_{i\rightarrow j})^\mu_{\;\nu}=g^{(i)\mu\lambda}g^{(j)}_{\lambda\nu}$, while the elementary symmetric polynomials can be explicitly constructed iteratively in terms of the trace of $S$, starting from $e_0(S)=1$, as:
\begin{equation}\label{sym pols}
    e_m(S)=-\frac{1}{m}\sum_{n=1}^m(-1)^n\Tr(S^n)e_{m-n}(S) \; .
\end{equation}
Lastly, since $S_{i\rightarrow j}=S_{j\rightarrow i}^{-1}$, there is a sense in which these interactions are \emph{oriented}: we say that a term in the potential, Eq. \eqref{MultigravPotential}, that contains $S_{i\rightarrow j}$ (\emph{not} $S_{j\rightarrow i}$) is positively oriented with respect to the $i$-th metric and negatively oriented with respect to the $j$-th metric. The orientation of an interaction with respect to a given metric affects the form of that metric's field equations, as we will see.

\begin{figure}[h!]
\centering
    \includegraphics[width=0.38\textwidth]{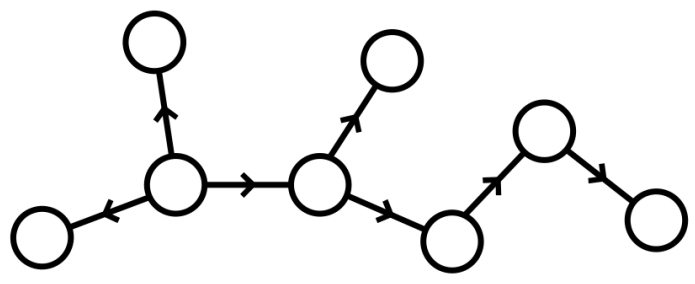}
    \caption{Directed theory graph representing some generic multi-metric model. The nodes represent different metrics, the edges indicate interactions and the arrows point in the direction of positive interaction orientation. Each metric generically has a number of interactions of either orientation, and each edge contributes a term to the field equations of the two metrics it connects; these terms are orientation-dependent.}
    \label{fig:interaction structure}
\end{figure}

The simplest way to view the interaction structure of a given model is as a directed graph \cite{SC_and_graph_structure,prop_bg_multigrav}, as depicted in figure \ref{fig:interaction structure}. Symmetries of a particular multi-metric model under permutations of the metric labels and swapping of interaction orientations can then be equivalently viewed as symmetries of said model's directed theory graph.

While somewhat unwieldy, it is crucial that the interactions between the metrics take this form, lest the Boulware-Deser ghost \cite{BD_Ghosts} be resurrected. Precisely, it is the Hamiltonian constraint within the theory that kills the ghostly degree of freedom; departing from the interaction structure given by Eq. \eqref{MultigravPotential} leads to this constraint becoming dynamical and so revives the ghost \cite{ghost_freedom_flat_ref,ghost_freedom_general_ref,ghost_freedom_stuckelberg,hamiltonian_analysis,Kluson,Kluson_note,Covariant_approach_no_ghosts}. The requirement of ghost freedom further restricts one to consider only those multi-metric theories that do not involve any interaction cycles (a cycle is e.g. $1\rightarrow2\rightarrow3\rightarrow1$, so the potential contains all three of $S_{1\rightarrow2}$, $S_{2\rightarrow3}$ and $S_{3\rightarrow1}$); the interactions between metrics must strictly form a tree graph \cite{ghost_freedom_multigravity,cycles}.

The vacuum field equations that arise from the action \eqref{MultigravAction} are as follows:
\begin{equation}\label{Einstein eqs}
    M^{D-2}_{i} \Gi + \Wi = 0 \; ,
\end{equation}
where the new term $W$ characterises the effect of the interactions over and above the standard GR interactions. In the metric formalism, it is explicitly given by:
\begin{equation}\label{W_metric}
\begin{split}
    \Wi &= \sum_j \sum_{m=0}^D(-1)^m\beta_m^{(i,j)}Y_{(m)\nu}^\mu(S_{i\rightarrow j})
    \\
    &+\sum_k\sum_{m=0}^D (-1)^m\beta_{D-m}^{(k,i)}Y_{(m)\nu}^\mu (S_{k\rightarrow i}^{-1}) \; ,
\end{split}
\end{equation}
where (with respect to the $i$-th metric) $j$ denote positively oriented interactions, $k$ denote negatively oriented interactions, and we define
\begin{equation}\label{Y_def}
    Y_{(m)}(S) = \sum_{n=0}^m (-1)^n S^{m-n}e_n(S) \; .
\end{equation}
As an example, to see how this works more explicitly, the metric furthest to the right in figure \ref{fig:interaction structure} has only a single negatively oriented interaction with its nearest neighbour. Denoting the furthest right metric as $g^{(I)}$ and the nearest neighbour as $g^{(J)}$, the vacuum field equations for $g^{(I)}$ read:
\begin{equation}
    M_I^{D-2} G^{(I)\mu}_{\;\;\;\;\;\;\nu} + \sum_{m=0}^D(-1)^m\beta_{D-m}^{(J,I)}Y^\mu_{(m)\nu}(S^{-1}_{J\rightarrow I}) = 0 \; .
\end{equation}

Finally, as a result of the Bianchi identities for each Einstein tensor, the covariant divergence of each $W$-tensor must also vanish:
\begin{equation}\label{W constraint}
    \nabla^{(i)\mu}W^{(i)}_{\mu\nu} = 0 \;\;\; \forall \, i \; .
\end{equation}
We refer to this condition as the \emph{Bianchi constraint}; it tells us that there can be no flow of energy-momentum across the interacting metrics.

\subsection{Black hole solutions: proportional, non-proportional and partially proportional branches}

The black hole solutions we constructed in part I can be subdivided into three distinct classes, depending on whether the various metrics are simultaneously diagonalisable \cite{BHs_multigrav}.

The cleanest and most unified way to construct these solutions is to work with the following Kerr-Schild ansatz for the multi-gravity metrics \cite{Gibbons_KdS,Rotating_AdS_bigravity}:
\begin{equation}\label{KdS}
   \gi = a_i^2\left(g^{(\Lambda)}_{\mu\nu} + \frac{r_{s,i}}{U} l_\mu l_\nu\right) \; ,
\end{equation}
where the $a_i$ are constant conformal factors\footnote{In fact, the Bianchi constraint enforces their constancy \cite{consistent_spin2}.} (one of which can be fixed by coordinate rescaling but the rest are physical as all metrics live on the same spacetime manifold), $g^{(\Lambda)}$ is the $D$-dimensional (A)dS metric with cosmological constant $\Lambda$, the $r_{s,i}$ are independent Schwarzschild radii for each metric, $U$ is some scalar function of the coordinates, and $l$ is a vector tangent to a null-geodesic congruence on (A)dS. We again direct the reader to part I for explicit expressions of these various functions \cite{BHs_multigrav}, as they are fairly long-winded and can differ depending on whether the spacetime dimension is odd or even. 

Continuing on with this ansatz, the Einstein tensors are diagonal, with components:
\begin{equation}\label{Einstein tens}
    \Gi = -\frac{\Lambda}{a_i^2}\delta^\mu_\nu \; ,
\end{equation}
while the $W$-tensor components become:
\begin{widetext}
\begin{equation}\label{W_nondiag}
    \begin{split}
        \Wi &= \delta^\mu_\nu \left[\sum_j\sum_{m=0}^D\beta_m^{(i,j)}\binom{D-1}{m}a_{j}^m a_i^{-m} + \sum_k\sum_{m=0}^D\beta_{D-m}^{(k,i)}\binom{D-1}{m}a_{k}^m a_i^{-m} \right]
        \\
        &+ \frac{l^\mu l_\nu}{2U} \left[\sum_j\frac{a_{j}}{a_i}\left(r_{s,i}-r_{s,j}\right)\sigma_{i,j}^{(+)} + \sum_k\frac{a_{k}}{a_i}\left(r_{s,i}-r_{s,k}\right)\sigma_{i,k}^{(-)}\right] \; ,
    \end{split}
\end{equation}
\end{widetext}
\clearpage \noindent where we have defined\footnote{The eagle-eyed reader may note that we have changed notation from part I slightly here, from $\Sigma_{i}^{(\pm)}\rightarrow\sigma_{i,j}^{(\pm)}$. This is to more readily account for general interaction structures with any number of postively and negatively oriented interactions per metric, as opposed to only accounting for the chain and star type interactions as we did originally in \cite{BHs_multigrav} (as well as to avoid clutter with the summation symbols).}:
\begin{align}
    \sigma^{(+)}_{i,j} &=  \sum_{m=0}^D \beta_m^{(i,j)}\binom{D-2}{m-1}a_{j}^{m-1}a_i^{1-m} \; ,\label{sig_p}
    \\
    \sigma^{(-)}_{i,k} &= \sum_{m=0}^D \beta_{D-m}^{(k,i)}\binom{D-2}{m-1}a_{k}^{m-1}a_i^{1-m} \; .\label{sig_m}
\end{align}
The plus and minus sigma variants are related to one another by:
\begin{equation}\label{Sig_pm}
     \sigma^{(-)}_{j,i} = \left(\frac{a_{i}}{a_{j}}\right)^{D-2} \sigma^{(+)}_{i,j} \; ,
\end{equation}
so we see that the $\sigma$'s live on the interaction links between any given pairs of metrics (since for an interaction built from $S_{i\rightarrow j}$, $\Wi$ will contain $\sigma_{i,j}^{(+)}$ while $W^{(j)\mu}_{\;\;\;\;\;\;\nu}$ will contain $\sigma_{j,i}^{(-)}$, but now we see these terms are just proportional to one another; put another way, if one vanishes, so does the other).

In order for the ansatz Eq. \eqref{KdS} to be a solution of the multi-metric field equations, two things must happen. First, the diagonal ($\delta^\mu_\nu$) part of the $W$-tensors must specify the value of the effective cosmological constant. More precisely, directly from the field equations one finds that the cosmological constant should satisfy the following set of $N$ simultaneous equations \cite{BHs_multigrav}:
\begin{equation}\label{Lambda_def}
\begin{split}
    \frac{\Lambda M_i^{D-2}}{a_i^2} &= \sum_j \sum_{m=0}^{D} \beta_m^{(i,j)} \binom{D-1}{m}a_{j}^m a_{i}^{-m} \\  &+\sum_k\sum_{m=0}^{D} \beta_{D-m}^{(k,i)} \binom{D-1}{m} a_{k}^m a_{i}^{-m} \; ,\;\;\;\forall\, i
\end{split}
\end{equation}
which, after fixing one of the $a_i$ via coordinate rescaling, can be solved exactly for $\Lambda$ and the remaining $N-1$ conformal factors, the physical solutions being those with real $\Lambda$ and $a_i$. 

Second, the off-diagonal ($l^\mu l_\nu$) components of the $W$-tensors must vanish. This can be achieved in multiple ways. The simplest is to take all of the Schwarzschild radii to be the same:
\begin{equation}\label{same_rs}
    r_{s,i}=r_s \;\;\; \forall \, i \; ;
\end{equation}
these are the \emph{proportional} solutions, since all the metrics become simultaneously diagonal, and are proportional to a common GR black hole solution.

The next simplest is to make all of the $\sigma$'s vanish:
\begin{equation}\label{Sig=0}
    \sigma_{i,j}^{(+)}=0 \;\;\;\forall\, i,j \; ;
\end{equation}
these are the \emph{non-proportional} solutions, since all of the Schwarzschild radii are independent and the metrics cannot be simultaneously diagonalised. For a given set of $\beta_m^{(i,j)}$ (which fully specify the multi-metric model in question) these are polynomial equations that fix the ratios of neighbouring conformal factors. To ensure that the field equations are satisfied, the same ratios that satisfy Eqs. \eqref{Sig=0} must \emph{also} satisfy Eqs. \eqref{Lambda_def}. This, however, is only true for certain choices of $\beta_m^{(i,j)}$. That is to say, if we assert a priori that $\sigma_{i,j}^{(+)}=0$, then only one of Eqs. \eqref{Lambda_def} actually fixes $\Lambda$, while the rest act as constraints on which multi-metric theories permit this class of solutions.

Finally, the most general solutions allow for combinations of both conditions \eqref{same_rs} and \eqref{Sig=0} for different interaction pairs; these are the \emph{partially proportional} solutions, since only some of the metrics share the same Schwarzschild radii and can be simultaneously diagonalised. In detail, if there are $n$ total interaction links that have $\sigma_{i,j}^{(+)}=0$ ($n$ can therefore be at most $N-1$), $n+1$ of the conformal factors are a priori fixed before checking whether the cosmological constant equations are satsified (the additional 1 is fixed by coordinate rescaling). Consequently, the $N$ diagonal equations \eqref{Lambda_def} split into $N-n$ algebraic equations for $\Lambda$ and the remaining $N-n-1$ free conformal factors, as well as $n$ equations that become constraints on the $\beta_m^{(i,j)}$ parameters of the theory. Again, this means that only for finely tuned interaction coefficients can these solutions exist.

In terms of the graph structure of figure \ref{fig:interaction structure}, each edge in a given theory graph corresponds to a specific interaction between a pair of metrics, and so comes equipped with its own $\sigma_{i,j}^{(+)}$ and its own pair of associated Schwarzchild radii (one for each of the two metrics/nodes adjoined by the edge in question). One may choose to either set $\sigma_{i,j}^{(+)}=0$ or $r_{s,i}=r_{s,j}$ along every edge in the theory graph; the three branches of solutions outlined above correspond to the different choices one can make in implementing this procedure. This diagrammatic picture also makes it clear that the partially proportional branch exists only for $N>2$ metrics: if one has exactly 2 metrics then the corresponding theory graph contains a single edge, so there is only a single choice to make in setting either $\sigma_{i,j}^{(+)}=0$ (non-proportional) or $r_{s,i}=r_{s,j}$ (proportional) along it.

The three branches of black hole solutions described above to date comprise all the analytically known black hole solutions of multi-metric gravity, and since they are based on the Kerr-Schild ansatz \eqref{KdS}, they contain the multi-metric analogues of all known black hole solutions of GR -- in part I we dubbed them ``GR-adjacent" black holes for this reason. Owing to the existence of non-GR-adjacent solutions in bigravity describing black holes supplemented by a cloud of massive graviton hair \cite{Hairy_BH_AdS,Hairy_BHs_flat,Hairy_BHs_Gervalle}, we further conjectured that similar hairy solutions should exist also in the full multi-metric theory, but omitted their determination due to the complexity of the calculation. As we mentioned in the introduction, we shall begin to cease shirking this responsibility this time around and give a procedure by which one may construct them generally in section \ref{Sec:hairy}, alongside an explicit calculation of the asymptotic form of the solution for a toy model with $N=3$ metrics, although we still save the full numerical calculation for part III. Before we get to that point, however, we would like to discuss the stability of the GR-adjacent solutions. Part I took care of the proportional branch, which we will recap below, then in section \ref{Sec:stability} we will extend these results to the non-proportional and partially proportional branches.

\subsection{Instability of the proportional branch}\label{sec:prop instability}

Perhaps the main result of part I was that the instabilities of the proportional branch of solutions which have been known for a decade in bigravity \cite{GL_instability_bigravity,Kerr_instability_bigravity,GL_unified} carry over naturally to the full multi-metric theory. To see this, one must linearise the field equations and consider the dynamics of the metric perturbations. Around the proportional solutions, which we can always write as $\gi=a_i^2\bar{g}_{\mu\nu}$ for some GR black hole solution $\bar{g}_{\mu\nu}$, this calculation simplifies greatly, as the perturbations acquire a standard Fierz-Pauli mass term: we showed in \cite{BHs_multigrav} that the spin-2 mass eigenstates $H^{(i)}_{\mu\nu}$ (which are linear combinations of the original metric perturbations) have dynamics given by:
\begin{equation}\label{TT equations}
    \bar{\Box}H^{(i)}_{\mu\nu} + 2\bar{R}^{\alpha\;\beta}_{\;\mu\;\nu}H^{(i)}_{\alpha\beta} = m_i^2 H^{(i)}_{\mu\nu} \; ,
\end{equation}
where the barred quantities are constructed from the common background metric $\bar{g}_{\mu\nu}$, and all indices are manipulated with $\bar{g}_{\mu\nu}$. The graviton square masses $m_i^2$ are the eigenvalues of the following mass matrix:
\begin{align}
    \mathcal{M}_{ii}^2 &= \frac{a_i^2}{M_i^{D-2}} \left(\sum_j \frac{a_{j}}{a_i}\sigma_{i,j}^{(+)} + \sum_k\frac{a_{k}}{a_i}\sigma_{i,k}^{(-)}\right) \label{Mii} \; ,
    \\
    \mathcal{M}_{ji}^2&=\left(\frac{a_{j}}{a_i}\right)^{4-D}\mathcal{M}_{ij}^2 = -\frac{a_i^2 \sigma_{i,j}^{(+)}}{\left(M_{j}M_i\right)^{\frac{D-2}{2}}}\label{Mi1} \; ,
\end{align}
which one can show always possesses one 0 eigenvalue \cite{ClockworkCosmo}, so the theory propagates at linear level a single massless graviton and $N-1$ massive gravitons.

Eqs. \eqref{TT equations} with $\bar{g}_{\mu\nu}$ taken as the $D=4$ Schwarzschild(-(A)dS) metric are precisely those equations studied in the context of the Gregory-Laflamme (GL) instability plaguing black string solutions in theories with extra dimensions \cite{GL_instability,Charged_GL_instability,AdS_GL_instability}. Therefore, the very same instability is present in the multi-metric theory too; precisely, it transpires that the proportional Schwarzschild(-(A)dS) solution becomes unstable if, for any of the $m_i$ \cite{BHs_multigrav,Kerr_instability_bigravity}:
\begin{equation}\label{GL instability}
    m_i r_s\lesssim0.876 \; .
\end{equation}

Put more plainly: a Schwarzschild black hole in any theory containing massive gravitons becomes unstable if the horizon size of the black hole in question becomes comparable to or smaller than (roughly) the Compton wavelength of the theory's lightest massive graviton. The form of this instability is exactly the same as the GL instability of a black string.

In part I we argued that this should be unsurprising: one may think of gravity in an extra compact dimension as the so-called `continuum limit' of a multi-metric theory with chain type interactions, where the various metrics are to be thought of as corresponding to discrete locations in the extra dimension, separated by some distance $\delta y$ \cite{discrete_grav_dims,deconstructing_dims,Deconstructing,ClockworkCosmo}. The continuum limit is taken by sending $N\rightarrow\infty$ and $\delta y\rightarrow0$ while keeping their product fixed (corresponding to the size of the additional compact dimension). The 5-dimensional black string solutions, which are well-known to suffer from the GL instability, are precisely what becomes of our proportional multi-metric black hole solutions in this limit, so it makes sense that the same behaviour should also be present in the discrete theory. 

The really interesting part is that the instability is present purely in the 4-dimensional multi-metric theory, even away from the continuum limit, and even for more generic interaction structures than the chain, suggesting that the GL instability may not be an inherently extra dimensional phenomenon, but may in fact be related to the fundamental nature of massive spin-2 interactions. Indeed, we would like to propose that the reason the GL instability is present in the black string system is not one of extra dimensional origin at all; rather, it is a consequence of the fact that the Fourier modes of the black string look like massive gravitons!

If $\bar{g}_{\mu\nu}$ is taken to be the $D=4$ Kerr metric, thus allowing for black hole rotation, then the GL instability is still present. The graviton mass threshold at which it becomes active is altered, but remains roughly at $m_i r_s\sim\mathcal{O}(1)$ \cite{backreaction}. There are also superradiant instabilities associated with the azimuthal modes that are not present in the Schwarzschild case \cite{Kerr_instability_bigravity,superradiance_ULS2,superradiance_massive_spin2,KG_rotating_BH,superradiance2020}, but it was shown in \cite{backreaction} that their growth is subdominant to that of the GL mode in the vast majority of the parameter space. Thus, a proper understanding of how the GL instability saturates in the multi-metric theory is necessary, which will require a full nonlinear analysis using numerical simulations (work is ongoing to make this a well-posed dynamical problem -- see \cite{Dynamical_dRGT} for the case of dRGT massive gravity with flat reference metric). Still, we shall see in section \ref{Sec:hairy} that there is a strong suggestion that the hairy solutions we provide a blueprint for may represent the true end state of the instability.

\section{Linear stability of non- and partially proportional solutions}\label{Sec:stability}

We turn now to look at the non- and partially proportional black hole solutions. In bigravity, with $N=2$ metrics, it has been known for some time that the non-proportional Schwarzschild solution is classically mode stable  \cite{Stability_nonbidiag,GL_unified}, owing to a rather unconventional, non-Fierz-Pauli form for the perturbation equations that leads to the black holes sharing their quasi-normal modes (QNMs) with those of the standard Schwarzschild solution in GR. As a result, only the two tensor modes of the massive graviton appear to propagate at linear level, while the vector and scalar modes are non-dynamical. We will have some more to say on this peculiar feature later; the goal initially, however, is to simply extend the bigravity analysis to the full multi-metric theory, as well as to the partially proportional branch, to complete our cataloguing of the linear stability of multi-metric black holes.

As mentioned, away from the proportional branch the structure of the perturbations becomes highly non-trivial, so for simplicity we shall work in this section exclusively in $D=4$, and restrict ourselves to non-rotating black holes only (namely, multi-Schwarzschild-(A)dS black holes). The spherical symmetry of these solutions will aid the stability analysis to follow, as we will see.

Following the approach taken in bigravity \cite{GL_unified,Stability_nonbidiag}, it proves useful to express the background metrics in (advanced) Eddington-Finkelstein coordinates as:
\begin{equation}\label{EF_coords}
        \dd \bar{s}^2_{(i)} = a_i^2\Bigg[-\left(1-\frac{r_{s,i}}{r}-\frac{\Lambda}{3}r^2\right)\dd v^2 + 2\dd v\dd r + r^2\dd\Omega_2^2 \Bigg] \; ,
\end{equation}
which is of course related to the Kerr-Schild form in Eq. \eqref{KdS} by a coordinate transformation. In these coordinates, the characteristic building block matrices take the simple form:
\begin{equation}
    S_{i\rightarrow j} =
    \begin{bmatrix}
        \frac{a_j}{a_i} & 0 & 0 & 0
        \\
        \frac{a_j}{a_i}\frac{(r_{s,j}-r_{s,i})}{2r} & \frac{a_j}{a_i} & 0 & 0
        \\
        0 & 0 & \frac{a_j}{a_i} & 0
        \\
        0 & 0 & 0 & \frac{a_j}{a_i}
    \end{bmatrix} \; .
\end{equation}
By substituting this into Eq. \eqref{Y_def} and then Eq. \eqref{W_metric}, one finds that the only non-vanishing off diagonal background $W$-tensor components are the $\bar{W}^{(i)r}_{\;\;\;\;\;v}$ terms, given by:
\begin{align}
    \bar{W}^{(i)r}_{\;\;\;\;\;v} = \frac{1}{2r} &\Bigg[\sum_j\frac{a_{j}}{a_i}\left(r_{s,i}-r_{s,j}\right)\sigma_{i,j}^{(+)} \nonumber
    \\
    &+ \sum_k\frac{a_{k}}{a_i}\left(r_{s,i}-r_{s,k}\right)\sigma_{i,k}^{(-)}\Bigg] \; ,
\end{align}
reflecting the form of Eq. \eqref{W_nondiag}.

We now perturb the metrics as:
\begin{equation}\label{perturbed mets}
    \gi = \bar{g}^{(i)}_{\mu\nu} + \delta g^{(i)}_{\mu\nu} \; .
\end{equation}
The spherical symmetry of the background allows one to decompose the perturbations into separate axial and polar contributions, expanded in a complete basis of tensor spherical harmonics. It also ensures that the various azimuthal modes of these contributions completely decouple from one another. In Fourier space, the decomposition reads:
\begin{equation}\label{ax_pol_decomposition}
    \delta\gi(v,r,\theta,\phi) = \sum_{l,m} \frac{a_i^2}{\sqrt{2\pi}} \int_{-\infty}^{\infty}\dd\omega\, e^{-\text{i}\omega v} \delta\tilde{g}^{(i)l m}_{\mu\nu}(\omega,r,\theta,\phi) \; ,
\end{equation}
where
\begin{equation}
    \delta\tilde{g}^{(i)lm}_{\mu\nu} = \delta\tilde{g}^{(i)\text{ax},lm}_{\mu\nu} + \delta\tilde{g}^{(i)\text{pol},lm}_{\mu\nu} \; ,
\end{equation}
with the axial and polar contributions explicitly given by (suppressing $(i)$ indices, and indicating symmetric components with asterisks) \cite{Kerr_instability_bigravity,Stability_nonbidiag}:
\begin{widetext}
\begin{align}
    \delta\tilde{g}^{\text{ax},lm}_{\mu\nu} &=\label{hax}
    \begin{bmatrix}
        0 & 0 & h_0^{lm}(\omega,r)\csc\theta\partial_\phi Y_{lm}(\theta,\phi) & - h_0^{lm}(\omega,r)\sin\theta\partial_\theta Y_{lm}(\theta,\phi)
        \\
        * & 0 & h_1^{lm}(\omega,r)\csc\theta\partial_\phi Y_{lm}(\theta,\phi) & - h_1^{lm}(\omega,r)\sin\theta\partial_\theta Y_{lm}(\theta,\phi)
        \\
        * & * & -h_2^{lm}(\omega,r)\csc\theta X_{lm}(\theta,\phi) & h_2^{lm}(\omega,r)\sin\theta Z_{lm}(\theta,\phi)
        \\
        * & * & * & h_2^{lm}(\omega,r)\sin\theta X_{lm}(\theta,\phi)
    \end{bmatrix} \; ,
    \\
    \delta\tilde{g}^{\text{pol},lm}_{\mu\nu} &=\label{hpol}
    \begin{bmatrix}
        H_0^{lm}(\omega,r)Y_{lm}(\theta,\phi) & H_1^{lm}(\omega,r)Y_{lm}(\theta,\phi) & \eta_0^{lm}(\omega,r)\partial_\theta Y_{lm}(\theta,\phi) & \eta_0^{lm}(\omega,r)\partial_\phi Y_{lm}(\theta,\phi)
        \\
        * & H_2^{lm}(\omega,r)Y_{lm}(\theta,\phi) & \eta_1^{lm}(\omega,r)\partial_\theta Y_{lm}(\theta,\phi) & \eta_1^{lm}(\omega,r)\partial_\phi Y_{lm}(\theta,\phi)
        \\
        * & * & r^2\Big[\begin{smallmatrix}K^{lm}(\omega,r)Y_{lm}(\theta,\phi)
        \\+G^{lm}(\omega,r)Z_{lm}(\theta,\phi)\end{smallmatrix}\Big] & r^2 G^{lm}(\omega,r) X_{lm}(\theta,\phi)
        \\
        * & * & * & r^2\sin^2\theta\Big[\begin{smallmatrix}K^{lm}(\omega,r)Y_{lm}(\theta,\phi) \\ - G^{lm}(\omega,r)Z_{lm}(\theta,\phi)\end{smallmatrix}\Big]
    \end{bmatrix} \; .
\end{align}
\end{widetext}
Here, $Y_{lm}(\theta,\phi)$ are the spherical harmonics, the functions $X_{lm}(\theta,\phi)$ and $Z_{lm}(\theta,\phi)$ are given by:

\begin{align}
    X_{lm}(\theta,\phi) &= 2\partial_\phi\left(\partial_\theta Y_{lm}-\cot\theta Y_{lm}\right) \; ,
    \\
    Z_{lm}(\theta,\phi) &= \partial_\theta^2 Y_{lm}-\cot\theta\partial_\theta Y_{lm} -\csc^2\theta\partial_\phi^2 Y_{lm} \; ,
\end{align}
while all of the functions of $(\omega,r)$ are free.

The full details of the procedure to extract the linearised field equations from metric perturbations around a generic background (i.e. not necessarily proportional) are outlined in appendix \ref{app:lin}, following the framework laid out originally in \cite{linear_spin2_gen_BG} for bigravity. The form of the contribution of a particular interaction to the linearised $W$-tensor is hugely dependent on which option of $r_{s,i}=r_{s,j}$ or $\sigma_{i,j}^{(+)}=0$ is chosen to make the off-diagonal part of the background $W$-tensor vanish along the associated link. 

As we saw in section \ref{sec:prop instability}, choosing the first (proportional) option, where neighbouring metrics share the same Schwarzschild radius, leads to a contribution of the standard Fierz-Pauli form; precisely:
\begin{equation}\label{lin W prop}
    \delta W^{(i)}_{\mu\nu} \supset \frac{\mathcal{M}^2_{ij}}{2}\left(\delta g^{(j)}_{\mu\nu} - \bar{g}^{(i)}_{\mu\nu} \delta g^{(j)}\right) \; .
\end{equation}

Choosing the second (non-proportional) option, however, leads instead to a highly non-trivial but remarkably simple contribution, which using the decomposition given by Eqs. \eqref{ax_pol_decomposition}, \eqref{hax}, \eqref{hpol} takes the form:
\begin{equation}\label{lin W nonprop}
    \delta\Wi \supset \frac{\mathcal{A}_{i,j}^{(\pm)}(r_{s,i}-r_{s,j})}{4r}e^{-\text{i}\omega v} (\Delta_{i,j})^\mu_{\;\nu} \; ,
\end{equation}
where the matrix $\Delta_{i,j}$ has components \cite{GL_unified,Stability_nonbidiag}:
\begin{widetext}
    \begin{equation}
        \Delta_{i,j} =
        \begin{bmatrix}
            0 & 0 & 0 & 0
            \\
            2[K^{lm}]_{i,j} Y_{lm} & 0 & -\Big(\begin{smallmatrix}
                \csc\theta\partial_\phi Y_{lm}[h_1^{lm}]_{i,j}
                \\+\partial_\theta Y_{lm}[\eta_1^{lm}]_{i,j}\end{smallmatrix}\Big) & \sin\theta\partial_\theta Y_{lm}[h_1^{lm}]_{i,j} - \partial_\phi Y_{lm}[\eta_1^{lm}]_{i,j}
            \\
            -\Big(\csc\theta\partial_\phi Y_{lm}[h_1^{lm}]_{i,j}+\partial_\theta Y_{lm}[\eta_1^{lm}]_{i,j}\Big) & 0 & [H_2^{lm}]_{i,j} Y_{lm} & 0
            \\
            \frac{1}{r^2\sin^2\theta}\Big(\partial_\theta Y_{lm}[h_1^{lm}]_{i,j}-\csc\theta\partial_\phi Y_{lm}[\eta_1^{lm}]_{i,j}\Big) & 0 & 0 & [H_2^{lm}]_{i,j} Y_{lm}
        \end{bmatrix} \; .
    \end{equation}
\end{widetext}
Here we have introduced the notation:
\begin{equation}
    [x]_{i,j} \equiv x^{(j)}-x^{(i)} \; ,
\end{equation}
to denote differences between the various perturbation functions of neighbouring metrics. Lastly, the constants out front are defined as:
\begin{align}
    \mathcal{A}_{i,j}^{(+)} &= 2\left(\frac{a_j^2}{a_i^2}\beta_2^{(i,j)} + \frac{a_j^3}{a_i^3}\beta_3^{(i,j)}\right) \; ,
    \\
    \mathcal{A}_{i,k}^{(-)} &= 2\left(\frac{a_k^2}{a_i^2}\beta_2^{(k,i)} + \frac{a_k}{a_i}\beta_3^{(k,i)}\right) \; ,
\end{align}
with the $(+)$/$(-)$ variants associated to positively/negatively oriented interactions. Note that $\beta_1$ is not present in these expressions, as we have used $\sigma_{i,j}^{(+)}=0$ to express it in terms of $\beta_2$ and $\beta_3$. Also note that in the bigravity case where one typically takes $a_0=1,a_1=C$ then the linearised $W$-tensors given above reduce precisely to the expressions quoted in \cite{Stability_nonbidiag}, as they should.

If we denote the whole expression on the right hand side of Eq. \eqref{lin W nonprop} by $\Xi_{i,j}^{(\pm)}$ i.e.
\begin{equation}
    \Xi_{i,j}^{(\pm)} = \frac{\mathcal{A}_{i,j}^{(\pm)}(r_{s,i}-r_{s,j})}{4r}e^{-\text{i}\omega v} \Delta_{i,j} \; ,
\end{equation}
then, like the $\sigma$'s, one finds that the $\Xi$'s are related by:
\begin{equation}\label{Xi_pm}
    \Xi_{i,j}^{(+)} = - \frac{a_j^4}{a_i^4} \Xi_{j,i}^{(-)} \; ,
\end{equation}
and so also live on the interaction links between pairs of metrics. 

In both the proportional and non-proportional cases, $\delta\Wi$ must satisfy the linearised version of the Bianchi constraint; that is:
\begin{equation}
    \nabla^{(i)}_\mu\delta\Wi = 0 \;\;\; \forall\,i \; .
\end{equation}
For proportional interactions this enforces the transverse-traceless gauge condition upon the mass eigenstates \cite{GL_instability_bigravity,BHs_multigrav}. For non-proportional interactions, each term contributes to the divergence as follows:
\begin{align}
    \nabla^{(i)}_\mu(\Xi_{i,j}^{(\pm)})^\mu_{\;\nu} &\propto\pm
    \begin{bmatrix}
        r\left(r[K^{lm}]_{i,j}\right)'+\frac{l(l+1)}{2}[\eta_1^{lm}]_{i,j}
        \\
        [H_2^{lm}]_{i,j}
        \\
        r[H_2^{lm}]_{i,j} - \left(r[\eta_1^{lm}]_{i,j}]\right)'
        \\
        \left(r[h_1^{lm}]_{i,j}\right)'\label{Bianchi lin nonprop}
    \end{bmatrix} \; .
\end{align}
These terms live on the interaction links too, owing to Eq. \eqref{Xi_pm}. Now we have all the information we need to discuss the stability of the remaining two branches of black hole solutions. We begin with the non-proportional branch.

\subsection{Non-proportional branch}\label{Sec:nonprop stability}

For the non-proportional solutions, every interaction has $\sigma_{i,j}^{(+)}=0$, so all contributions to the $W$-tensors are of the form \eqref{lin W nonprop}. Explicitly, the linearised field equations are:
\begin{equation}
    \mathcal{E}^{(i)\alpha\beta}_{\;\;\;\;\,\mu\nu}\delta g^{(i)}_{\alpha\beta} + \Lambda \delta\gi + \frac{1}{M_i^2}\left[\sum_j \Xi_{i,j}^{(+)}+\sum_k\Xi_{i,k}^{(-)}\right] = 0 \; ,
\end{equation}
where the Lichnerowicz operator $\mathcal{E}^{(i)\alpha\beta}_{\;\;\;\;\,\mu\nu}$ is defined in appendix \ref{app:lin}, while the linearised Bianchi constraints are:
\begin{equation}
    \sum_j \nabla^{(i)}_\mu(\Xi_{i,j}^{(+)})^\mu_{\;\nu} + \sum_k \nabla^{(i)}_\mu(\Xi_{i,k}^{(-)})^\mu_{\;\nu} = 0 \; .
\end{equation}

\begin{figure}[h!]
\centering
    \includegraphics[width=0.38\textwidth]{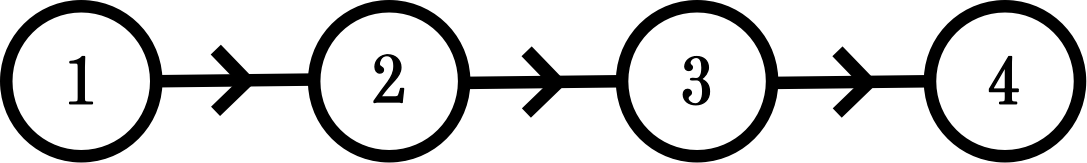}
    \caption{Chain theory with 4 metrics and all interactions positively oriented. Each edge (interaction) contributes a term to the Bianchi constraint of each of the metrics it connects, and these terms are proportional to one another (C.F. Eq. \eqref{Xi_pm}). The Bianchi constraint on metric 1 implies the vanishing of the term coming from the $1\rightarrow2$ edge, and the Bianchi contraint on metric 4 implies the vanishing of the term coming from the $3\rightarrow4$ edge. Together with the Bianchi constraints on metrics 2 and 3, these then imply the vanishing of the term coming from the $2\rightarrow3$ edge.}
    \label{fig:4 metric chain}
\end{figure}

At first glance, the sums over $j$ and $k$, within which the number of terms can differ from metric to metric, appear to make the Bianchi constraints awkward to resolve. Thankfully, this is not the case. To see why, one may consider all of the metrics within the interaction structure that possess only a single interaction (there must always be at least 2 such metrics). The Bianchi constraints for these metrics contain sums over $j$ or $k$ that have only one term, which must therefore vanish on its own. However, since these terms live on interaction links, there is always a proportional term of opposite orientation in the neighbouring metric's Bianchi constraint, which must also vanish, thus reducing the number of terms present in said constraint by one. This process propagates through the entirety of the interaction structure until every metric ends up with only one term left in its Bianchi constraint, which of course must then also vanish; the result is that \emph{every} term in the sums over $j$ and $k$ must vanish \emph{individually}. To convince oneself that this must happen, consider a simple example with just 4 metrics interacting in a chain structure -- see figure \ref{fig:4 metric chain}.

The 4 Bianchi constraints read:
\begin{align}
    \nabla^{(1)}_\mu(\Xi_{1,2}^{(+)})^\mu_{\;\nu} &= 0\label{nab0} \; ,
    \\
    \nabla^{(2)}_\mu(\Xi_{2,1}^{(-)})^\mu_{\;\nu} + \nabla^{(2)}_\mu(\Xi_{2,3}^{(+)})^\mu_{\;\nu} &= 0\label{nab1} \; ,
    \\
    \nabla^{(3)}_\mu(\Xi_{3,2}^{(-)})^\mu_{\;\nu} + \nabla^{(3)}_\mu(\Xi_{3,4}^{(+)})^\mu_{\;\nu} &= 0\label{nab2} \; ,
    \\
    \nabla^{(4)}_\mu(\Xi_{4,3}^{(-)})^\mu_{\;\nu} &= 0\label{nab3} \; .
\end{align}
The vanishing of Eqs. \eqref{nab0} and \eqref{nab3} imply the vanishing of the first term in Eq. \eqref{nab1} and the second term in Eq. \eqref{nab2}; which then implies the vanishing of the final remaining divergence.

From Eq. \eqref{Bianchi lin nonprop}, the vanishing of every term in the sums over $j$ and $k$ can immediately be resolved to find the following expressions for the differences in perturbation functions:
\begin{align}
    [H_2^{lm}]_{i,j} &= 0\label{H2} \; ,
    \\
    [\eta_1^{lm}]_{i,j} &= \frac{C_{i,j}}{r}\label{eta1} \; ,
    \\
    [h_1^{lm}]_{i,j} &= \frac{D_{i,j}}{r}\label{h1} \; ,
    \\
    [K^{lm}]_{i,j} &= \frac{E_{i,j}}{r} + \frac{C_{i,j}l(l+1)}{2r^2} \; ,\label{K}
\end{align}
for all $i,j$, and where the $C,D$ and $E$ are arbitrary integration constants. Since these hold across all interactions, one may express all of the perturbation functions in terms of those of a single distinguished metric (e.g. one may express all other $\eta_1^{(i)lm}$ in terms of $\eta_1^{(1)lm}$ simply by using Eq. \eqref{eta1} repeatedly).

\subsubsection{Quasi-normal modes}

To see that these results imply mode stability at the linear level, one must consider the QNMs, as was shown in bigravity \cite{Stability_nonbidiag}. These are the eigenvalues of the linearised field equations with suitable boundary conditions: namely, that the perturbations behave as ingoing waves near the black hole horizon and as outgoing waves at infinity, {\it viz}.
\begin{equation}\label{QNM_BCs}
    \delta\tilde{g}^{(i)}_{\mu\nu} \rightarrow A^{(i)\pm}_{\mu\nu} e^{\pm k_{\pm} r_{*i}} \; .
\end{equation}
Here, the plus sign refers to the boundary condition as $r\rightarrow\infty$, the minus sign refers to the boundary condition as $r\rightarrow r_{s,i}$, $A_{\mu\nu}^{(i)\pm}$ are typically polynomials in $1/r$, and $r_{*i}$ are the tortoise coordinates for each metric, defined by $\dd r/\dd r_{*i} = (1-r_{s,i}/r)$.

By simple inspection of Eqs. \eqref{H2}--\eqref{K} together with the decomposition \eqref{ax_pol_decomposition} into axial and polar contributions, one finds that it is impossible to satisfy these boundary conditions unless \emph{all} of the $C_{i,j}=D_{i,j}=E_{i,j}=0$ \cite{Stability_nonbidiag}. This is easy to see: consider, for example, the $r\phi$-components; one has:
\begin{equation}
    [\delta\tilde{g}^{lm}_{r\phi}]_{i,j} = e^{-\text{i}\omega r_*}\left(\frac{C_{i,j}}{r}\partial_\phi Y_{lm}-\frac{D_{i,j}}{r}\sin\theta\partial_\theta Y_{lm}\right) \; ,
\end{equation}
which represents an ingoing wave of frequency $\omega$ (C.F. Eq. \eqref{ax_pol_decomposition}), and the same must be true of $\delta\tilde{g}^{(i)lm}_{r\phi}$ and $\delta\tilde{g}^{(j)lm}_{r\phi}$ individually. The near-horizon boundary condition is therefore always satisfied, while the near-infinity boundary condition never can be, unless the integration constants vanish. The same is true for all components, and of course for all interactions, since as we mentioned earlier, the perturbation functions of any given metric may be expressed in terms of those of, say, $\tilde{g}^{(i)lm}_{\mu\nu}$.

Therefore, for the non-proportional solutions, \emph{all} of the perturbation matrices $\Delta_{i,j}$ completely vanish, and the linearised field equations reduce to just $N$ copies of the standard linearised GR equations:
\begin{equation}
    \mathcal{E}^{(i)\alpha\beta}_{\;\;\;\;\,\mu\nu}\delta g^{(i)}_{\alpha\beta} + \Lambda \delta\gi = 0 \; .
\end{equation}
As a result, the eigenvalue problem defined by the multi-metric field equations with boundary conditions \eqref{QNM_BCs} is equivalent to that of $N$ Schwarzschild(-(A)dS) metrics in standard GR, just where each metric now has a different Schwarzschild radius $r_{s,i}$.

We already know that the Schwarzschild(-(A)dS) solution in GR has no unstable QNMs (and that its QNMs only exist at all for $l\geq2$ -- there is no monopole nor dipole), so the same is true here: the non-proportional multi-metric solutions are classically mode stable, thus generalising the result from bigravity. This behaviour is in stark contrast to that of the proportional solutions, whose QNM spectrum is very different from the Schwarzschild solution in GR, as it depends on the graviton masses and contains more dynamical modes (one of which -- the radial $l=0$ mode -- is unstable, as we have seen).

\subsubsection{Generic gravitational perturbations}

Although the QNMs of these non-proportional black hole solutions coincide with those of a Schwarzschild black hole in GR and so exhibit no modal instability, strictly speaking, the absence of unstable modes in the quasi-normal spectrum of a black hole does not guarantee full stability of the solution. In principle, one should also relax the QNM boundary condition at infinity, to further allow for ingoing waves, as this tracks the linear response of the spacetime to perturbations wrought by external sources. 

The full set of linear gravitational perturbations, upon relaxing this boundary condition, were worked out for bigravity in \cite{Stability_nonbidiag}. While, for both our non-proportional solutions and the standard Schwarzschild solution in GR, the QNM spectra contain only modes with multipoles $l\geq2$, in principle the full set of perturbations, including those caused by external perturbers, may also contain dynamical modes with $l=0$ and $l=1$. In GR, these modes of the Schwarzschild metric are pure gauge and are hence non-dynamical. In contrast, the authors of \cite{Stability_nonbidiag} showed that for the non-proportional black holes in bigravity, there are in fact new physical modes that exist in both the axial and polar sectors that cannot be gauged away. However, oddly, these new modes feel no effective potential and so are not backscattered at all by the black hole geometry (a property more reminscent of Minkowski spacetime); they describe purely ingoing waves whose propagation is unaltered throughout the entire space. 

Since the structure of the perturbation equations is the same in the full multi-metric theory, the same qualitative behaviour occurs in the generic non-proportional solutions here too, beyond bigravity to theories involving arbitrary numbers of interacting metrics. This is due to the fact that Eqs. \eqref{H2}--\eqref{K} mean that the perturbation functions of all the different metrics still take the same functional form, up to arbitrary multiplicative constants.

\subsubsection{Nonlinear instability?}

All of this is a little strange: we have found that around the non-proportional solutions, only 2 of the 5 degrees of freedom (the tensor modes) of each of the massive gravitons actually propagate at linear level, and that modes exist in the complete spectrum of gravitational perturbations that feel no effective potential, despite the (apparent) presence of a black hole. Consequently, one may begin to worry about the possibility of strong coupling, or relatedly, whether problems arise when the interactions of the missing degrees of freedom reappear at nonlinear level. We touched upon this potentially troublesome feature very briefly in part I. We will now be more bold as to make a conjecture. First, some motivation: the condition on the model parameters that gives rise to the non-proportional branch of black hole solutions, $\sigma_{i,j}^{(+)}=0$, also rears its head in the realm of multi-metric cosmology. There, FLRW solutions are also split into three distinct branches, depending this time not on how to make off-diagonal $W$-tensor components vanish, but on how one chooses to satisfy the Bianchi constraint along each interaction link \cite{ClockworkCosmo,FRW_cosmology_dRGT,self_accelerating_branch_1,self_accelerating_branch_2}. The analogue of the proportional branch involves fixing the lapses as functions of the scale factors, while the analogue of the non- and partially proportional branches requires one to take precisely the same condition on the parameters of the model, $\sigma_{i,j}^{(+)}=0$, across all or some of the interaction links. Considering perturbations around the `non-proportional' branch of cosmological solutions, one finds exactly the same behaviour that we find for the analogous black hole solutions -- namely, that the vector and scalar degrees of freedom are not dynamical at linear level in the perturbations \cite{mg_cosmo_perts,perts_open_FRW}. However, after taking nonlinear effects into account, these cosmological solutions turn out to be unstable, owing to the emergence of (non-BD) ghosts as the vector and scalar degrees of freedom reenter the effective description \cite{nonlinear_instability_FRW,nonlinear_cosmo_stability,general_mass,viable_cosmo_bigravity}. Therefore, it is natural to conjecture that the same thing will happen for the non-proportional black hole solutions: that, ultimately, this branch will prove to be ghostly, and consequently pathological. Proving this conjecture may be difficult -- in cosmology, it was done by considering \emph{linear} perturbations around an anisotropic Bianchi-I background as a proxy for nonlinear perturbations around the isotropic FLRW background \cite{nonlinear_cosmo_stability}. For the black hole solutions, one presumably must go genuinely nonlinear, to at least quadratic order in metric perturbations. We will not perform this calculation here, but we have good reason to expect that the result will be the same as in the cosmological case.

Similarly, if one were to consider perturbations around a non-proprtional Kerr background, thus allowing for black hole rotation, the calculation of the linearised field equations for all modes becomes tricky because there is no longer any spherical symmetry to allow one to decompose the perturbations as in Eq. \eqref{ax_pol_decomposition}. Nevertheless, one expects that at linear level these non-proprtional rotating solutions will share their QNMs with a standard Kerr black hole in GR, but will become unstable nonlinearly.

Another potential sign of a pathology in the non-proportional branch is that one may send signals from inside to outside a black hole. For example, suppose we have two metrics, $g^{(1)}$ and $g^{(2)}$, with horizons at $r_{s,1}$ and $r_{s,2}<r_{s,1}$ respectively, with matter (minimally) coupled to $g^{(1)}$. An observer, who is made of matter, sees the horizon $r_{s,1}$ and when they fall through it they may send a signal by generating gravity waves of $g^{(1)}$, before they reach $r_{s,2}$. However, $g^{(1)}$ and $g^{(2)}$ are coupled, so this means that gravity waves of $g^{(2)}$ are created, which are outside their horizon, $r_{s,2}$, and so may freely propagate to infinity, whereupon an observer may detect them through their interaction with $g^{(1)}$. From the perspective of $g^{(2)}$, there is a source of waves that is travelling outside the light cone of $g^{(2)}$, and so will source Cherenkov radiation. We expect the backreaction of this radiation to source an instability of the geometry.

\subsection{Partially proportional branch}\label{Sec:partial prop stability}

To understand what happens for the partially proportional solutions, it helps to begin with a simple example. We will again choose to consider the 4 metric chain model depicted in figure \ref{fig:4 metric chain}, and work with the particular partially proportional solution where metrics 2, 3 and 4 are proportional to one another but metric 1 is not (one finds this solution by choosing $\sigma_{1,2}^{(+)}=0$ along the $1\rightarrow2$ edge, and choosing the Schwarzschild radii to be the same along the remaining two edges). This means that the contribution of the $1\rightarrow2$ interaction to the $W$-tensors is of the form \eqref{lin W nonprop}, while the contributions of the $2\rightarrow3$ and $3\rightarrow4$ interactions to the $W$-tensors are of the form \eqref{lin W prop}. Explicitly, the linearised field equations here read:
\begin{widetext}
    \begin{align}
        \mathcal{E}^{(1)\alpha\beta}_{\;\;\;\;\,\mu\nu}\delta g^{(1)}_{\alpha\beta} + \Lambda \delta g^{(1)}_{\mu\nu} + \frac{\Xi_{1,2}^{(+)}}{M_1^2} &= 0 \; ,\label{linmet1}
        \\
        \mathcal{E}^{(2)\alpha\beta}_{\;\;\;\;\,\mu\nu}\delta g^{(2)}_{\alpha\beta} + \Lambda \delta g^{(2)}_{\mu\nu} + \frac{\Xi_{2,1}^{(-)}}{M_2^2} + \frac{\mathcal{M}^2_{22}}{2}\left(\delta g^{(2)}_{\mu\nu} - {g}^{(2)}_{\mu\nu} \delta g^{(2)}\right) + \frac{\mathcal{M}^2_{23}}{2}\left(\delta g^{(3)}_{\mu\nu} - {g}^{(2)}_{\mu\nu} \delta g^{(3)}\right) &= 0 \; ,\label{linmet2}
        \\
        \mathcal{E}^{(2)\alpha\beta}_{\;\;\;\;\,\mu\nu}\delta g^{(3)}_{\alpha\beta} + \Lambda \delta g^{(3)}_{\mu\nu} + \frac{\mathcal{M}^2_{32}}{2}\left(\delta g^{(2)}_{\mu\nu} - {g}^{(2)}_{\mu\nu} \delta g^{(2)}\right) + \frac{\mathcal{M}^2_{33}}{2}\left(\delta g^{(3)}_{\mu\nu} - {g}^{(2)}_{\mu\nu} \delta g^{(3)}\right) + \frac{\mathcal{M}^2_{34}}{2}\left(\delta g^{(4)}_{\mu\nu} - {g}^{(2)}_{\mu\nu} \delta g^{(4)}\right) &= 0 \; ,\label{linmet3}
        \\
        \mathcal{E}^{(2)\alpha\beta}_{\;\;\;\;\,\mu\nu}\delta g^{(4)}_{\alpha\beta} + \Lambda \delta g^{(4)}_{\mu\nu} + \frac{\mathcal{M}^2_{43}}{2}\left(\delta g^{(3)}_{\mu\nu} - {g}^{(2)}_{\mu\nu} \delta g^{(3)}\right) + \frac{\mathcal{M}^2_{44}}{2}\left(\delta g^{(4)}_{\mu\nu} - {g}^{(2)}_{\mu\nu} \delta g^{(4)}\right) &= 0 \; ,\label{linmet4}
    \end{align}
\end{widetext}
where in the latter 3 equations we consider the perturbations as living in the common background of $g^{(2)}_{\mu\nu}$, since metrics 3 and 4 are both proportional to $g^{(2)}_{\mu\nu}$. As before, to talk about stability one must turn to the QNMs, and we just showed that in order for the solutions to Eqs. \eqref{linmet1}--\eqref{linmet4} to satisfy QNM boundary conditions this implies that the contribution from the non-proportional interaction, $\Xi_{1,2}^{(+)}$, must vanish. Therefore, Eq. \eqref{linmet1} reduces to the field equations of linearised GR for metric 1, while the $\Xi_{2,1}^{(-)}$ term drops out of Eq. \eqref{linmet2} for metric 2, leaving only the contributions from the Fierz-Pauli terms coming from the proportional metrics. 

\begin{figure}[h!]
\centering
    \includegraphics[width=0.4\textwidth]{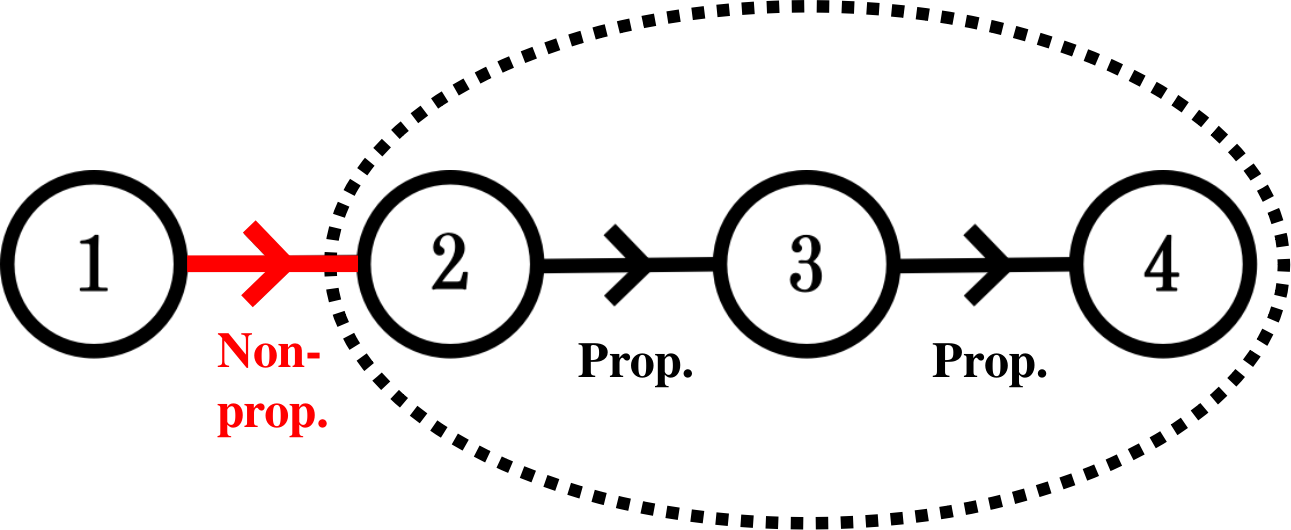}
    \caption{A partially proportional solution of the 4-metric chain theory introduced in figure \ref{fig:4 metric chain}. Since $\Xi_{1,2}^{(+)}=0$ for the non-proportional interaction along the $1\rightarrow2$ edge, at linear level metric 1 completely decouples from the remaining proportional metrics, whose linearised field equations are now equivalent to those one would get by linearising around a fully proportional solution of the tri-metric chain theory contained within the dashed oval.}
    \label{fig:partial prop 4 metric}
\end{figure}

Thus, the non-proportional part of the interaction structure completely decouples from the proportional part at linear level; what remains is a proportional \emph{sector} of the complete 4-metric theory, containing only metrics 2, 3 and 4, whose linearised field equations are the same as those that would be derived from a completely proportional black hole solution in a theory which contained only these 3 metrics in the first place (i.e. chain trigravity, without metric 1). Figure \ref{fig:partial prop 4 metric} shows this diagrammatically.

However, since it only accounts for the 3 proportional metrics, $\mathcal{M}^2$ no longer represents the true mass matrix of the full 4-metric theory i.e. its eigenvalues are \emph{not} the physical graviton masses. Nevertheless, its components are precisely as in Eqs. \eqref{Mii} and \eqref{Mi1} and so it does act as a sort of `local mass matrix' for the proportional sector -- indeed, if one was to look at the linearised field equations around proportional black holes in chain trigravity, rather than this partially proportional solution in our 4-metric chain theory, then $\mathcal{M}^2$ would be the physical mass matrix.

What certainly is true, as we found in part I and recapped briefly in section \ref{sec:prop instability} of this work, is that the equations in the proportional sector can exhibit an instability \emph{à la} Gregory-Laflamme, depending on the sizes of the eigenvalues of $\mathcal{M}^2$ relative to $r_{s,2}$ -- see Eq. \eqref{GL instability}. So, the partially proportional black hole solutions in multi-metric gravity also possess this instability, precisely because it still exists in the remaining proportional sectors. Around the fully proportional solutions, we could interpret these eigenvalues as the genuine physical masses of the various gravitons, but now, around the partially proportional solutions, they are simply numbers that tell us whether a given solution is unstable.

\begin{figure}[h!]
\centering
    \includegraphics[width=0.48\textwidth]{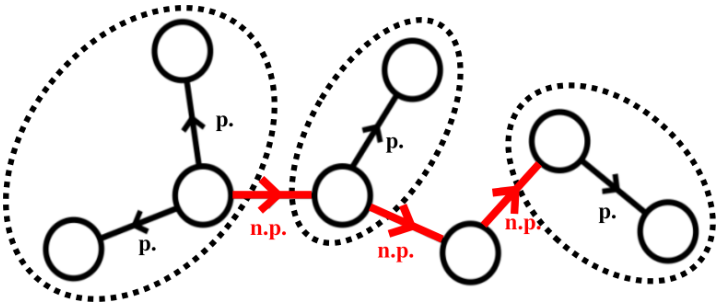}
    \caption{Example partially proportional solution in some generic multi-metric theory. This particular example contains 3 distinct proportional sectors indicated by the dashed ovals: there is an effective trigravity and two different effective bigravities. The solution is unstable if the eigenvalues of \emph{any} of the 3 local mass matrices satsisfy $m_i r_{s,i}\lesssim0.876$.}
    \label{fig:partial prop general}
\end{figure}

From this simple 4-metric example it is clear how the argument generalises: a partially proportional solution in some generic multi-metric theory will contain a number of both proportional (same $r_s$) and non-proportional ($\sigma=0$) interactions. The linearised Bianchi constraints, together with QNM boundary conditions, force all of the contributions of the non-proportional interactions to the various $W$-tensors to vanish. Thus, one is left with (in principle) multiple proportional sectors whose linearised field equations are the same as those of a fully proportional solution in a multi-metric theory containing the same number of metrics as the sector in question, while the remaining metrics with only non-proportional interactions follow the standard linearised GR equations. Each of the proportional sectors comes with its own `local mass matrix', and the eigenvalues of each of these matrices inform the conditions upon which the partially proportional solution exhibits the GL instability\footnote{Note that every one of these matrices has a 0 eigenvalue, so there are also as many `local massless modes' as there are proportional sectors.}. Figure \ref{fig:partial prop general} makes this more transparent.

Of course, the discussion up to this point is based on a purely linear analysis; if our conjecture about the non-proportional solutions becoming unstable at nonlinear level proves to be correct, then these partially proportional solutions will suffer the same fate because of the surviving \emph{non-proportional} sectors. This potential nonlinear instability is worse than the GL instability in the proportional sectors, as it is related to the emergence of ghostly degrees of freedom, and so represents an instability of the vacuum. Conversely, the GL instability simply indicates that the metric perturbations undergo exponential growth and so must backreact on the solution: it may be that the GL instability evolves into some other, stable, final state. As a result, we believe that the only really physically sensible GR-adjacent solutions in multi-metric gravity that might describe real black holes are the proportional solutions (which is handy, because they are the simplest) in the regime where there is no GL instability. As for what might happen after these solutions eventually turn unstable, we continue into the next section.

\section{Hairy black hole solutions}\label{Sec:hairy}

As a matter of necessity, the fact that the proportional black hole solutions in multi-gravity suffer from the GL instability means that they have to decay into something (provided that they actually form in the first place -- though if they do not, we have just seen that this probably means the theory has no viable black hole solutions). For the proportional multi-Schwarzschild solution, the final state should remain spherically symmetric, lest cosmic censorship be violated. It was this point that prompted the authors of \cite{Hairy_BH_AdS,Hairy_BHs_flat,Hairy_BHs_Gervalle} to search for other spherically symmetric solutions in bigravity, where they indeed discovered additional solutions bifurcating from the Schwarzschild one at precisely the point at which the GL instability switches on, describing black holes supplmented by massive graviton hair. It is important to stress that this was not confirmation that the proportional Schwarzschild solutions in bigravity genuinely decay into these hairy solutions. To find out for sure, one would need to use numerical relativity simulations, which do not even exist yet for bigravity in a well-posed form, never mind for the full multi-metric theory. However, it is still a strong suggestion that this may be the case, and it certainly motivates us to try to construct similar hairy solutions in the full multi-metric theory.

To that end, we consider the most general static and spherically symmetric ansatz we can use for the multi-gravity metrics:
\begin{equation}\label{hairy metrics}
    \dd s_{(i)}^2 = -p_i^2(r)\dd t^2 + \frac{U_i'^{2}(r)}{Y_i^2(r)}\dd r^2 + U_i^2(r)\dd\Omega_2^2 \; ,
\end{equation}
where the $p_i$, $Y_i$ and $U_i$ are radial functions. There is enough gauge freedom to allow us to fix \emph{one} of the $U_i(r)=r$, but the rest of the functions must remain free. Therefore, for $N$ metrics, there are $3N-1$ free functions to solve for. One may additionally try to include off-diagonal components in the metrics, but one finds that the only solutions that exist turn out to be the non- or partially proportional solutions we have already constructed \cite{Hairy_BH_AdS}, so we stick with the diagonal ansatz \eqref{hairy metrics} going forward.

The field equations and Bianchi constraints that result from this metric ansatz are as follows (see appendix \ref{app:hairy} for the explicit derivation):
\begin{widetext}
\begin{align}
    2Y_i^2U_iY_i' + Y_i U_i'\left[\left(Y_i^2-1\right) + A_{i,0}^{(+)}+A_{i,0}^{(-)}\right] + Y_i\left(\sum_j U_j'A_{i,j}^{(+)} + \sum_k U_k'A_{i,k}^{(-)}\right) &= 0 \; ,\label{hairy G00}
    \\\nonumber \\
    2Y_i^2 U_i p_i' + U_i'\left[p_i\left(Y_i^2-1\right) + B_{i,0}^{(+)}+B_{i,0}^{(-)}\right] &= 0 \; ,\label{hairy Grr}
    \\\nonumber \\
    U_i Y_i U_i'Y_i'p_i' + Y_ip_iU_i'^{2}Y_i'+Y_i^2\left[p_i''U_i'U_i + p_i'\left(U_i'^{2}-U_iU_i''\right)\right]\nonumber \\+ U_i'^3\left(C_{i,0}^{(+)}+C_{i,0}^{(-)}\right) + U_i'^2\left(\sum_j U_j'C_{i,j}^{(+)} + \sum_k U_k'C_{i,k}^{(-)}\right) &= 0 \; ,\label{hairy Gthth}
    \\\nonumber \\
    \sum_j \left[U_i'p_j'D_{i,j}^{(+)}-U_j'\left(p_i' \frac{Y_i}{Y_j}D_{i,j}^{(+)} + U_i'\left(\frac{Y_i}{Y_j}-1\right)E_{i,j}^{(+)}\right)\right]\nonumber \\+ \sum_k \left[U_i'p_k'D_{i,k}^{(-)}-U_k'\left(p_i' \frac{Y_i}{Y_k}D_{i,k}^{(-)} + U_i'\left(\frac{Y_i}{Y_k}-1\right)E_{i,k}^{(-)}\right)\right] &=0 \; ,\label{hairy Bianchi}
\end{align}
\end{widetext}
where the various functions $A_{i,j}^{(\pm)}$, $B_{i,j}^{(\pm)}$, $C_{i,j}^{(\pm)}$, $D_{i,j}^{(\pm)}$ and $E_{i,j}^{(\pm)}$ are functions of the $U$'s and $p$'s whose form is given explicitly in appendix \ref{app:hairy}. We note that Eqs. \eqref{hairy Bianchi} are the Bianchi constraints, which we know live on interaction links, so each term in the sums over $j$ and $k$ here must vanish individually, by the same argument as in section \ref{Sec:nonprop stability}.

One can check that by choosing 
\begin{align}
    p_i(r)&=a_i\sqrt{1-\frac{r_s}{r} -\frac{\Lambda}{3}r^2} \; ,
    \\
    Y_i(r)&=\frac{p_i(r)}{a_i} \; ,
    \\
    U_i(r)&=a_i r \; ,
\end{align}
the field equations reduce down to those of the proportional multi-Schwarzschild(-(A)dS) solution, with $\Lambda$ given by Eq. \eqref{Lambda_def}, as they should.

Solving these equations generally for a given multi-metric model is \emph{extremely} difficult, and it must be done numerically -- it is no surprise that the calculation has only been performed so far in bigravity. Nevertheless, it can in principle be done in the full multi-metric theory, as we demonstrate below. The idea is to treat the first derivatives of the various free functions initially as variables, in order to reduce the system into a set of coupled first order nonlinear ODEs, which can then be integrated up from the black hole horizon to infinity to determine the solution. To see that the system is solvable, we first consider the distinguished metric $g^{(i*)}_{\mu\nu}$, upon which we have used gauge freedom to fix $U_{i*}=r$. Eq. \eqref{hairy Grr} then allows one to find $p_{i*}'$ in terms of $r$ and the other free functions. There are now $2(N-1)$ remaining free $U_i'$ and $p_i'$ that are not yet isolated. Eqs. \eqref{hairy Bianchi} and \eqref{hairy Grr} for the remaining metrics are $2(N-1)$ equations in exactly these variables, so one may solve them as simultaneous equations for the remaining $U_i'$ and $p_i'$. Finally, one may substitute the now determined $U_i'$ into Eqs. \eqref{hairy G00} to obtain equations for the $Y_i'$, which then closes the system, and one can integrate this mass of coupled ODEs to find a solution. Note that Eqs. \eqref{hairy Gthth} are redundant in this whole procedure.

In practice, this process is incredibly tedious, and is complicated by the fact that the right hand sides of all the field equations are 0 -- this means that the determinant of the (in principle huge) matrix defined by the simultaneous equations for $p_i'$ and $U_i'$ must vanish for a solution to exist where the $p_i'$ and $U_i'$ are not simply 0 (which is obviously unphysical). This determinant must of course be resolved, and there are similar potential stumbling blocks at various stages of the calculation. We will choose to work with a simpler example model with just 3 metrics, which is already complicated, but nevertheless still contains the relevant physics.

\subsection{Example model: star trigravity}

\begin{figure}[h!]
\centering
    \includegraphics[width=0.33\textwidth]{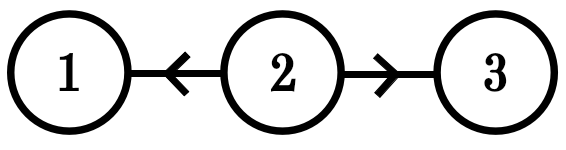}
    \caption{Theory graph for star trigravity. The theory is invariant under the exchange of metrics 1 and 3 (remember, a symmetry of the theory is a symmetry of its directed graph).}
    \label{fig:3 metric star}
\end{figure}

As a tractable example, we will consider the 3-metric star theory which takes the same gravitational coupling $M_i=M$ on all three sites, and also the same set of interaction coefficients for both interactions, $\beta_m^{(1,2)}=\beta_m^{(2,3)}=\beta_m$, specified by:
\begin{align*}
    \beta_0 &= -6M^2 m^2 \\
    \beta_1 &= 3M^2 m^2 \\
    \beta_2 &= -M^2 m^2 \\
    \beta_3&=\beta_4=0 \; ,
\end{align*}
introducing the new parameter $m$ which is related to the masses of the gravitons in a manner we specify below. These $\beta$'s are chosen so that there is a solution where all 3 metrics are equal to the standard Schwarzschild metric (i.e. with $\Lambda=0$ and where the proportionality factors are all $a_i=1$), and because the graviton masses around the proportional solutions can be determined analytically \cite{BHs_multigrav} to be $m_i=\{0,m,\sqrt{3}m\}$ -- we will see why this becomes important later. The star structure is also preferred over the chain here because the theory is symmetric under the exchange of metrics 1 and 3, which turns out to make the field equations slightly simpler to deal with. The model's theory graph is displayed in figure \ref{fig:3 metric star}.

We will choose to use the gauge freedom to set on the central metric $U_2(r)=r$; the free functions we must solve for are then $p_1$, $p_2$, $p_3$, $Y_1$, $Y_2$, $Y_3$, $U_1$ and $U_3$. The procedure to reduce the system to a set of first order ODEs is outlined below; the complete form of many of the expressions are extremely lengthy and uninstructive, though we make the derivation publically available in a Mathematica notebook online if one wishes to look at them and/or follow the derivation \cite{Zenodo}.
\begin{enumerate}
    \itemsep0em

    \item First, look at Eq. \eqref{hairy Grr} on metric 3. Solve it for $p_2'$; the solution looks like $p_2'=p_1(\hdots)+p_3(\hdots) +p_2(\hdots)$.

    \item Substitute this into Eq. \eqref{hairy Bianchi} on metric 3. Together with \eqref{hairy Grr} on metric 3 these are a pair of simultaneous equations for $p_3'$ and $U_3'$ both equalling 0. The determinant of the matrix defined by these equations must vanish for a non-trivial solution, which gives a relation of the form $p_3=p_2(\hdots)+p_1(\hdots)$.

    \item Do the same thing as in step 2 but now for metric 1 instead of metric 3. One gets a pair of simultaneous equations for $p_1'$ and $U_1'$ whose determinant must vanish, which gives $p_1=p_2(\hdots)+p_3(\hdots)$. 

    \item Steps 2 and 3 can be combined and resolved to write:
    \begin{align}
        p_1 &=p_2 G_1(r,U_1,U_3,Y_1,Y_2,Y_3)\label{G1}
        \\
        p_3 &=p_2 G_3(r,U_1,U_3,Y_1,Y_2,Y_3)\label{G3}
    \end{align}
    so the explicit $p$-dependence of the $(\hdots)$ functions has been extracted.

    \item Substitute Eqs. \eqref{G1} and \eqref{G3} into step 1 to find:
    \begin{equation}
        p_2' = p_2 F_2(r,U_1,U_3,Y_1,Y_2,Y_3)\label{F2}
    \end{equation}
    then solve Eq. \eqref{hairy Grr} on metrics 1 and 3 for $p_{1,3}'=(p_2(\hdots) + p_{1,3}(\hdots))U_{1,3}'$ and substitute in Eqs. \eqref{G1} and \eqref{G3} to get:
    \begin{align}
        p_1'&=p_2 F_1(r,U_1,U_3,Y_1,Y_2,Y_3)U_1'\label{F1}
        \\
        p_3'&=p_2 F_3(r,U_1,U_3,Y_1,Y_2,Y_3)U_3'\label{F3}
    \end{align}

    \item Use Eq. \eqref{hairy G00} on all 3 metrics to write $Y_1'=(\hdots)U_1'+(\hdots)$, $Y_2'=(\hdots)U_1'+(\hdots)U_3'+(\hdots)$ and $Y_3'=(\hdots)U_3'+(\hdots)$.

    \item Together, Eqs. \eqref{G1}--\eqref{F3} tell us that:
    \begin{align}
        G_1' &= F_1U_1'-F_2G_1
        \\
        G_3' &= F_3U_3'-F_2G_3
    \end{align}
    Taking the derivatives of $G_1$ and $G_3$ explicitly, and substituting in the expressions from step 6, gives 2 simultaneous equations for $U_1'$ and $U_3'$. Solve them.

    \item Substitute the solutions for $U_1'$ and $U_3'$ into step 6, and we finally get to a closed system of 5 coupled first-order nonlinear ODEs for the $Y$'s and $U$'s:
    \begin{align}
        Y_1' &= \mathcal{F}_1(r,U_1,U_3,Y_1,Y_2,Y_3)\label{Y1}
        \\
        Y_2' &= \mathcal{F}_2(r,U_1,U_3,Y_1,Y_2,Y_3)\label{Y2}
        \\
        Y_3' &= \mathcal{F}_3(r,U_1,U_3,Y_1,Y_2,Y_3)\label{Y3}
        \\
        U_1' &= \mathcal{F}_4(r,U_1,U_3,Y_1,Y_2,Y_3)\label{U1}
        \\
        U_3' &= \mathcal{F}_5(r,U_1,U_3,Y_1,Y_2,Y_3)\label{U3}
    \end{align}

    \item Solve (numerically) Eqs. \eqref{Y1}--\eqref{U3} for the $Y$'s and $U$'s. Substitute the solution into Eq. \eqref{F2} and integrate to find $p_2$. Substitute this into Eqs. \eqref{G1} and \eqref{G3} to determine the remaining free functions $p_1$ and $p_3$, then we are done.
    
\end{enumerate}

The same procedure will work in principle for any generic multi-metric theory beyond just the 3-metric star. The tedious part arises due to the fact that whenever there are additional metrics one is necessarily forced to propagate the Bianchi constraints through the whole interaction structure during steps 2--4, in a similar manner to how we did in section \ref{Sec:nonprop stability}, but ultimately one should eventually be able to arrive at equations of the form \eqref{G1}--\eqref{F2} for the $p$'s, and equations of the form \eqref{Y1}--\eqref{U3} for the $Y$'s and $U$'s, which can be integrated numerically. In practice, we of course do not recommend actually doing this; it is tricky enough with just 3 metrics. Indeed, we save the full numerical calculation for this example model for the final paper of this series. Despite the inherent complexity of the field equations, it turns out that one can still glean some useful physical information by looking at the asymptotic behaviour of these equations far away from the black hole horizon, which we now turn to.

\subsubsection{Asymptotic behaviour at infinity}

For the model specified by our choice of $\beta_m$, as $r\rightarrow\infty$, the spacetime becomes approximately Minkowski on all three metrics. Therefore, we shall consider the following asymptotic form of the free functions:
\begin{align}
    p_i(r) &= 1 + \delta p_i(r) \; ,
    \\
    Y_i(r) &= 1 + \delta Y_i(r) \; ,
    \\
    U_i(r) &= r + \delta U_i(r) \; ,
\end{align}
where the variations are small. Substituting these expansions into Eqs. \eqref{Y1}--\eqref{U3} and keeping only the linear terms leads to a remarkable simplification; one finds:
\begin{align}
    \delta Y_1' &= m^2\delta U_1 - \frac{1}{3r}\left(\delta Y_1 + \delta Y_2 + \delta Y_3\right) \; ,\label{deltaY1}
    \\
    \delta Y_2' &= -m^2\left(\delta U_1 + \delta U_3\right) - \frac{1}{3r}\left(\delta Y_1 + \delta Y_2 + \delta Y_3\right) \; ,\label{deltaY2}
    \\
    \delta Y_3' &= m^2\delta U_3 - \frac{1}{3r}\left(\delta Y_1 + \delta Y_2 + \delta Y_3\right) \; ,\label{deltaY3}
    \\
    \delta U_1' &= \left(1+\frac{4}{3m^2 r^2}\right)\delta Y_1 - \left(1+\frac{2}{3m^2r^2}\right)\delta Y_2\nonumber \\ &- \frac{2}{3m^2r^2}\delta Y_3 \; , \label{deltaU1}
    \\
    \delta U_3' &= -\frac{2}{3m^2r^2}\delta Y_1 - \left(1+\frac{2}{3m^2r^2}\right)\delta Y_2\nonumber \\ &+ \left(1+\frac{4}{3m^2 r^2}\right)\delta Y_3 \; . \label{deltaU3}
\end{align}
These equations are symmetric under $1\leftrightarrow3$ exchange, as we expect from the star interaction structure. They can be solved analytically to find:
\begin{widetext}
    \begin{align}
        \delta Y_1 &= -\frac{A}{2r} + \frac{B}{r}\left(1+mr\right)e^{-mr} + \frac{C}{r} \left(1+\sqrt{3}mr\right)e^{-\sqrt{3}mr}\label{dY1} \; ,
        \\
        \delta Y_2 &= -\frac{A}{2r} - \frac{2C}{r} \left(1+\sqrt{3}mr\right)e^{-\sqrt{3}mr}\label{dY2} \; ,
        \\
        \delta Y_3 &= -\frac{A}{2r} - \frac{B}{r}\left(1+mr\right)e^{-mr} + \frac{C}{r} \left(1+\sqrt{3}mr\right)e^{-\sqrt{3}mr}\label{dY3} \; ,
        \\
        \delta U_1 &= -\frac{B}{m^2r^2}\left(1+mr+m^2r^2\right)e^{-mr} - \frac{C}{m^2r^2}\left(1+\sqrt{3}mr+3m^2r^2\right)e^{-\sqrt{3}mr} \; , \label{dU1}
        \\
        \delta U_3 &= \frac{B}{m^2r^2}\left(1+mr+m^2r^2\right)e^{-mr} - \frac{C}{m^2r^2}\left(1+\sqrt{3}mr+3m^2r^2\right)e^{-\sqrt{3}mr} \; . \label{dU3}
    \end{align}
\end{widetext}
In principle there are also exponentially growing terms, but we have set to 0 their associated integration constants since their presence would spoil the asymptotic flatness of the solutions. Accounting for this fact numerically requires a deft hand \cite{Hairy_BHs_Gervalle}; we will give more details in part III.

Lastly, the equations for the $\delta p$'s linearise to:
\begin{align}
    \delta p_2' &= \frac{1}{3r}\left(\delta Y_1 + \delta Y_3 - 5\delta Y_2\right) \; ,
    \\
    \delta p_1 &= \delta p_2 - \frac{2}{3m^2 r^2} \left(3m^2 r \delta U_1 + 2\delta Y_1 - \delta Y_2 - \delta Y_3\right) \; ,
    \\
    \delta p_3 &= \delta p_2 - \frac{2}{3m^2 r^2} \left(3m^2 r \delta U_3 + 2\delta Y_3 - \delta Y_2 - \delta Y_1\right) \; .
\end{align}
Again these equations are symmetric under $1\leftrightarrow3$ exchange as we expect. Substituting in the solutions we have just obtained for the $\delta Y$'s and $\delta U$'s, they imply:
\begin{align}
    \delta p_1 &= -\frac{A}{2r} + \frac{2B}{r}e^{-mr} + \frac{2C}{r}e^{-\sqrt{3}mr} \; ,\label{dp1}
    \\
    \delta p_2 &= -\frac{A}{2r} - \frac{4C}{r}e^{-\sqrt{3}mr} \; ,\label{dp2}
    \\
    \delta p_3 &= -\frac{A}{2r} - \frac{2B}{r}e^{-mr} + \frac{2C}{r}e^{-\sqrt{3}mr} \; .\label{dp3}
\end{align}

So, we have now solved for all of the asymptotic functions: the $\delta U$'s are given by Eqs. \eqref{dU1} and \eqref{dU3}, the $\delta Y$'s by Eqs. \eqref{dY1}--\eqref{dY3}, and the $\delta p$'s by Eqs. \eqref{dp1}--\eqref{dp3}. The physical meaning of each of the terms in these expressions is also clear: the $A/r$ piece is the standard Newtonian potential coming from the massless graviton, but there are also Yukawa pieces coming from two massive gravitons whose masses are $m$ and $\sqrt{3}m$ -- precisely the masses we expected from diagonalising the mass matrix. The manner in which the superposition of these different effects appears in the form of the metric functions is not a trivial matter and clearly depends on the field equations, which are complicated (e.g. metric 2 doesn't see the graviton of mass $m$ in either its $\delta p$ or $\delta Y$ functions, and it appears that the sum of the pieces coming from the massive gravitons across all three $\delta p$'s or $\delta Y$'s must vanish -- this happened in bigravity too \cite{Hairy_BH_AdS,Hairy_BHs_flat,Hairy_BHs_Gervalle}). However, the behaviour is qualitatively sensible, and the same should be true in any more generic multi-metric theory: that is, the asymptotic behaviour will always consist of a massless Newtonian piece together with some superposition of Yukawa pieces corresponding to gravitons whose masses one may determine by diagonalising the mass matrix around a particular proportional solution. Explicitly, we expect that the result will be of the form:
\begin{align}
    \delta Y_i &= -\frac{A}{2r} + \sum_{j=1}^{N-1} \frac{B_{i,j}}{r}\left(1+m_j r\right)e^{-m_j r} \; ,
    \\
    \delta U_i &= \sum_{j=1}^{N-1} \frac{C_{i,j}}{m_j^2 r^2}\left(1+m_j r+ m_j^2 r^2\right)e^{-m_j r} \; ,
    \\
    \delta p_i &= - \frac{A}{2r} + \sum_{j=1}^{N-1} \frac{D_{i,j}}{r} e^{-m_j r} \; ,
\end{align}
where the $C_{i,j}$ and $D_{i,j}$ are constants related by simple rescalings to the arbitrary $B_{i,j}$, and one has for any given $j=J$, $\sum_i B_{i,J} = \sum_{i} D_{i,j} = 0$. Our results here for 3 metrics are certainly of this form, as are the bigravity results of \cite{Hairy_BHs_flat,Hairy_BH_AdS,Hairy_BHs_Gervalle}; we expect the same to be true in general.

The structure of these asymptotic solutions as a superposition of graviton mass modes is the reason that we interpret the solutions as describing black holes endowed with massive graviton hair. In order to determine the full solution, one must numerically integrate the complete set of coupled first order nonlinear ODEs from the horizon to infinity and ensure that the initial conditions are chosen such that the asymptotic behaviour given above is recovered. The asymptotics define a boundary value problem for the set of coupled ODEs, but they are only known up to arbitrary integration constants which we must somehow fix. The calculation is a challenging one, so we save it and its results for the eventual final paper of this series. Nevertheless, we have still given a general procedure one may follow if one wishes to construct hairy black hole solutions in generic multi-metric models.

\section{Conclusion}\label{Sec:conclusion}

In this work, we sought to build upon what we did previously in part I \cite{BHs_multigrav} by determining the linear stability of the non-proportional and partially proportional branches of black hole solutions in the general theory of ghost free multi-metric gravity, generalising and extending analogous results from dRGT massive gravity and bigravity. We showed that, as is the case in bigravity, the non-proportional multi-Schwarzschild solutions in the full multi-metric theory (at least in $D=4$) are mode stable at linear level, as each metric shares its quasi-normal spectrum with the standard Schwarzschild solution in GR. On the other hand, the partially proportional solutions (which only exist for $N>2$ metrics) are unstable at linear level, as they inherit the Gregory-Laflamme instability from the sectors of the solution in question that do still remain proportional. 

One should not be so hasty to accept these findings as statements of the full stability of the solutions, however, as linear stability does not always guarantee nonlinear stability. Indeed, the conclusions from our linear analysis may, and likely do, change at nonlinear level, as one expects the scalar and vector graviton helicity states -- that do not propagate at linear level around the non-proprtional solutions, and the non-proportional sectors of the partially proportional solutions -- to reappear nonlinearly and give rise to ghost-like instabilities. Such behaviour is known to occur already in the analogous branch of cosmological multi-metric solutions, where the disappearance of the vector and scalar degrees of freedom at linear level occurs similarly to here, so it is natural to conjecture that the nonlinear ghost instability might also arise for these classes of black hole solutions.

The consequence of this would be that the only physically sensible GR-adjacent black hole solutions in multi-metric gravity are the proportional solutions, at least in the regime where these solutions are stable. However, we also know from part I that even the proportional solutions become unstable when the black hole horizon size drops below the Compton wavelength of the theory's lightest (massive) graviton. All is not lost though -- the GL instability in the proportional branch is not necessarily as fatal as the aforementioned ghost instability in the non-proportional and partially proportional branches, as it only signifies an exponentially growing metric perturbation (rather than a genuine instability of the vacuum), which may backreact on the solution to eventually saturate into a new, stable final state.

To that end, we also considered a generic spherically symmetric ansatz for the multi-gravity metrics and showed how one may construct solutions describing black holes endowed with massive graviton hair. These solutions are expected to bifurcate from the multi-Schwarzschild one at the point at which it becomes unstable and so provide a good candidate for the end state of the instability. We considered an example model involving 3 metrics that is tractable enough to see how the procedure to construct hairy solutions works in practice. Although we save the full numerical methodology and calculation for part III, we showed how to reduce the system to a set of coupled nonlinear first order ODEs amenable to numerical integration, and analytically determined the asymptotic form of their solution far away from the black hole horizon, where one clearly sees the contributions from each of the individual graviton mass modes. Our plan for the final installment is to develop the numerical technology to fully integrate these equations from the horizon in order to determine the complete solution, and investigate how the resulting hairy solutions depend on the multi-gravity parameter space and on the masses of the gravitons. 

Obviously there is still much to learn about all of these black hole solutions, not least whether they genuinely do form from gravitational collapse in these multi-metric theories, and whether the GL instability in the proportional branch genuinely does evolve to one of the hairy solutions. These are questions which require numerical relativity simulations to answer with certainty -- we again stress the importance of developing a well-posed dynamical formulation of multi-metric gravity for this and many other purposes. That said, we have in this series of papers constructed a complete cataloguing of all the known black hole solutions, including their linear stability properties, of any generic multi-metric theory of gravity. We hope that this will prove invaluable to anyone wishing to study black holes in theories involving massive spin-2 fields.

\section*{Acknowledgements}

K.W. is supported by a UK Science and Technology Facilities Council studentship. P.M.S. and A.A. are supported by a STFC Consolidated Grant [Grant No. ST/T000732/1]. Some calculations involving the linearised $W$-tensors and hairy solutions were aided by use of the \emph{xAct} Mathematica package suite \cite{xAct} (specifically \emph{xCoba}). For the purpose of open access, the authors have applied a Creative Commons Attribution (CC BY) licence to any Author Accepted Manuscript version arising.

\section*{Data Access Statement}
No new data were created or analysed in this study, though we have made publically available the Mathematica notebook where we reduced the field equations for the 3-metric star model introduced in section \ref{Sec:hairy} down to a set of coupled ODEs, found at \cite{Zenodo}.

\appendix
\section{Linearised field equations around generic backgrounds}\label{app:lin}

Around the proportional backgrounds, the linearised field equations are not overly difficult to compute; indeed, we showed how to do this in appendix B of part I \cite{BHs_multigrav}. Around generic backgrounds, the calculation is much more complicated, but it can still be done in a systematic manner. The prescription for doing so was first laid out in detail in \cite{linear_spin2_gen_BG}, we outline it here to show how we determined the structure of the perturbations around the non-proportional solutions.

We begin by perturbing the metrics as in Eq. \eqref{perturbed mets}. It is well known (see e.g. \cite{dR_review}) that the Einstein tensor linearises to the Lichnerowicz operator acting on the perturbation, that is:
\begin{equation}
    \delta G^{(i)}_{\mu\nu} = \mathcal{E}^{(i)\alpha\beta}_{\;\;\;\;\,\mu\nu}\delta g^{(i)}_{\alpha\beta} \; ,
\end{equation}
where \cite{GL_instability_bigravity},
\begin{equation}\label{Lichnerowicz}
\begin{split}
    \mathcal{E}^{(i)\alpha\beta}_{\;\;\;\;\,\mu\nu}h_{\alpha\beta} = \frac12\bigl[ &-\Box^{(i)} h_{\mu\nu} + \nabla_\mu^{(i)} \nabla_\alpha^{(i)} h^\alpha_{\;\nu}  \\ &+ \nabla_\nu^{(i)} \nabla_\alpha^{(i)} h^\alpha_{\;\mu} - \nabla_\mu^{(i)} \nabla_\nu^{(i)} h
    + g_{\mu\nu}^{(i)} \Box^{(i)} h  \\ &- g_{\mu\nu}^{(i)}\nabla_\alpha^{(i)} \nabla_\beta^{(i)} h^{\alpha\beta} - 2R^{(i)\alpha\;\beta}_{\;\;\;\;\;\;\mu\;\nu}h_{\alpha\beta}\bigr] \; ,
\end{split}
\end{equation}
so, as in part I, we shall skip this part of the derivation here and focus on the potential. To that end, the first order variation of the $W$-tensor is:
\begin{equation}\label{delta W}
\begin{split}
        \delta\Wi &= \delta g^{(i)\mu}_{\;\;\;\;\;\lambda}\bar{W}^{(i)\lambda}_{\;\;\;\;\;\;\nu} 
        \\
        &+ \sum_j \sum_{m=0}^D(-1)^m\beta_m^{(i,j)}\delta Y_{(m)\nu}^\mu(S_{i\rightarrow j})
        \\
        &+ \sum_k\sum_{m=0}^D(-1)^m\beta_{D-m}^{(k,i)}\delta Y_{(m)\nu}^\mu(S_{i\rightarrow k}) \; .
\end{split}
\end{equation}
The variation of the $Y$'s is given (in matrix notation, and for any given $i\rightarrow j$ interaction) by:
\begin{align}\label{delta Y}
    \delta Y_{(m)}(S) = \sum_{k=1}^m (-1)^k &\Bigg[ S^{m-k}\delta e_k(S)
    \\
    &-e_{k-1}(S)\sum_{n=0}^{m-k} S^n \delta S S^{m-k-n} \Bigg] \; ,\nonumber
\end{align}
where, by virtue of Eq. \eqref{sym pols}, we have:
\begin{equation}\label{delta ek}
    \delta e_k(S) = -\sum_{n=1}^k (-1)^n\Tr(S^{n-1}\delta S) e_{k-n}(S) \; .
\end{equation}

The complication lies in the fact that $\delta S$ is given by the \emph{matrix} equation:
\begin{equation}\label{Sylvester eq}
    S\delta S + \delta S S = \delta S^2 \; ,
\end{equation}
and so one cannot simply determine it by starting from $S^2_{i\rightarrow j}=g^{-1}_{(i)}g_{(j)}$ and then Taylor expanding the square root (unless $S\propto\mathbbm{1}$, which \emph{is} the case for the proportional solutions, and is the reason that they are simpler to deal with). Around a generic background solution, this is a \emph{Sylvester matrix equation}, of the form:
\begin{equation}
    AX-XB=C \; ,
\end{equation}
where $A$, $B$ and $C$ are given constant matrices and one wishes to solve for the unknown matrix $X$. The solution to the Sylvester equation is known in the mathematical literature, and is given by the following expression \cite{Sylvester_matrix_eq}:
\begin{equation}
    X = q_B^{-1}(A)\sum_{k=1}^D\sum_{n=0}^{k-1}(-1)^k e_{D-k}(B)A^{k-n-1}CB^n \; ,
\end{equation}
where $q_B(A)$ is the unique polynomial in the matrix $A$ whose coefficients are the same as those of the characteristic polynomial of $B$ ($q_B^{-1}(A)$ is then the inverse of this matrix); that is:
\begin{equation}
    q_B(A) = \sum_{m=0}^D (-1)^m e_{D-m}(B) A^m \; .
\end{equation}

In our case, comparison with Eq. \eqref{Sylvester eq} tells us that we have $A=S$, $B=-S$ and $C=\delta S^2$. Therefore, the solution for $\delta S$ is:
\begin{equation}\label{delta S}
    \delta S = q^{-1}_{-S}(S)\sum_{k=1}^D\sum_{n=0}^{k-1}(-1)^{n+k} e_{D-k}(-S)S^{k-n-1}\delta S^2 S^n \; .
\end{equation}
One can easily obtain $\delta S^2_{i\rightarrow j}$ in terms of either the metric perturbations of $g_{(i)}$ and $g_{(j)}$, or of their inverses, by starting from $S^2_{i\rightarrow j}=g^{-1}_{(i)}g_{(j)}$ and substituting in Eq. \eqref{perturbed mets} for the perturbed metrics. The result is:
\begin{align}
    \delta S^2_{i\rightarrow j} &= g_{(i)}^{-1}\left[\delta g_{(j)}-\delta g_{(i)}S^2_{i\rightarrow j}\right] 
    \\
    &= \left[S^2_{i\rightarrow j}\delta g_{(j)}^{-1}-\delta g_{(i)}^{-1}\right]g_{(j)} \; ,
\end{align}
or in components:
\begin{align}
    (\delta S^2_{i\rightarrow j})^\mu_{\;\nu} &= g_{(i)}^{\mu\lambda}\left[\delta g_{(j)\lambda\nu}-\delta g_{(i)\lambda\sigma}(S^2_{i\rightarrow j})^\sigma_{\;\nu}\right] 
    \\
    &= \left[(S^2_{i\rightarrow j})^\mu_{\;\lambda}\delta g_{(j)}^{\lambda\sigma}-\delta g_{(i)}^{\mu\sigma}\right]g_{(j)\sigma\nu} \; .
\end{align}

Substituting either of these expressions into Eq. \eqref{delta S} determines $\delta S$, which one can then substitute into Eq. \eqref{delta Y} to get the $Y$ variations, and lastly substitute these into Eq. \eqref{delta W} to determine the linearised $W$-tensors. Around the non-proportional black hole solutions, one finds that the linearised $W$-tensors take the form of Eq. \eqref{lin W nonprop}.

\section{Background field equations for generic spherically symmetric diagonal metrics}\label{app:hairy}

With the ansatz Eq. \eqref{hairy metrics} for the metrics, the Einstein tensor components are:
\begin{align}
    G^{(i)0}_{\;\;\;\;\;0} &= \frac{Y_i^2}{U_i'^2}\left[2\frac{Y_i'}{Y_i}\frac{U_i'}{U_i}+\frac{U_i'^2}{Y_i^2 U_i^2}\left(Y_i^2-1\right)\right]
    \\ \nonumber \\
    G^{(i)r}_{\;\;\;\;\;r} &= \frac{Y_i^2}{U_i'^2}\left[2\frac{p_i'}{p_i}\frac{U_i'}{U_i}+\frac{U_i'^2}{Y_i^2 U_i^2}\left(Y_i^2-1\right)\right]
    \\ \nonumber \\
    G^{(i)\theta}_{\;\;\;\;\;\theta} &= G^{(i)\phi}_{\;\;\;\;\;\phi} \nonumber
    \\
    &= \frac{Y_i^2}{U_i'^2}\left[\frac{Y_i'}{Y_i}\left(\frac{U_i'}{U_i}+\frac{p_i'}{p_i}\right) + \frac{p_i''}{p_i}+ \frac{p_i'}{p_i}\left(\frac{U_i'}{U_i}-\frac{U_i''}{U_i}\right)\right]
\end{align}

For \emph{positively} oriented interactions, the contributions to the $W$-tensors are given by:
\begin{widetext}
    \begin{align}
        W^{(i)0}_{\;\;\;\;\;\;0} &\supset \left(\beta_0^{(i,j)} + 2\beta_1^{(i,j)}\frac{U_j}{U_i} + \beta_2^{(i,j)}\frac{U_j^2}{U_i^2}\right) + \frac{Y_i U_j'}{Y_j U_i'}\left(\beta_1^{(i,j)} + 2\beta_2^{(i,j)}\frac{U_j}{U_i} + \beta_3^{(i,j)}\frac{U_j^2}{U_i^2}\right)
        \\
        W^{(i)r}_{\;\;\;\;\;\;r} &\supset \left(\beta_0^{(i,j)}+\beta_1^{(i,j)}\frac{p_j}{p_i}\right)+ 2\left(\beta_1^{(i,j)}+\beta_2^{(i,j)}\frac{p_j}{p_i}\right)\frac{U_j}{U_i} + \left(\beta_2^{(i,j)}+\beta_3^{(i,j)}\frac{p_j}{p_i}\right)\frac{U_j^2}{U_i^2}
        \\
        W^{(i)\theta}_{\;\;\;\;\;\;\theta}&=W^{(i)\phi}_{\;\;\;\;\;\;\phi}\nonumber
        \\
        &\supset \left(\beta_0^{(i,j)}+\beta_1^{(i,j)}\frac{U_j}{U_i}\right) + \left(\beta_1^{(i,j)}+\beta_2^{(i,j)}\frac{U_j}{U_i}\right)\frac{p_j}{p_i} + \frac{Y_i U_j'}{Y_j U_i'}\left[\left(\beta_1^{(i,j)}+\beta_2^{(i,j)}\frac{U_j}{U_i}\right) + \left(\beta_2^{(i,j)}+\beta_3^{(i,j)}\frac{U_j}{U_i}\right)\frac{p_j}{p_i}\right]
    \end{align}
\end{widetext}
For \emph{negatively} oriented interactions, one should replace in the above expressions $j\rightarrow k$ and $\beta_m^{(i,j)}\rightarrow\beta_{4-m}^{(k,i)}$.

The Bianchi constraints have only one non-vanishing contribution from each interaction, which is (written for a postively oriented interaction -- again make the exchanges $j\rightarrow k$ and $\beta_m^{(i,j)}\rightarrow\beta_{4-m}^{(k,i)}$ to obtain the appropriate expression for a negatively oriented inteaction):
\begin{widetext}
    \begin{equation}
    \begin{split}
        \nabla^{(i)}_\mu &W^{(i)\mu}_{\;\;\;\;\;\;r} \supset \left(\beta_1^{(i,j)} + 2\beta_2^{(i,j)}\frac{U_j}{U_i} + \beta_3^{(i,j)}\frac{U_j^2}{U_i^2}\right)\frac{p_j'}{p_i} 
        \\ 
        &- \frac{Y_i U_j'}{Y_j U_i'}\left\{\left(\beta_1^{(i,j)} + 2\beta_2^{(i,j)}\frac{U_j}{U_i} + \beta_3^{(i,j)}\frac{U_j^2}{U_i^2}\right)\frac{p_i'}{p_i} + 2\left[\left(\beta_1^{(i,j)}+\beta_2^{(i,j)}\frac{U_j}{U_i}\right) + \left(\beta_2^{(i,j)}+\beta_3^{(i,j)}\frac{U_j}{U_i}\right)\frac{p_j}{p_i}\right]\left(1-\frac{Y_j}{Y_i}\right)\frac{U_i'}{U_i}\right\}
    \end{split}
    \end{equation}
\end{widetext}

Substituting the above expressions into the field equations $M_i^2\Gi + \Wi=0$, and simplifying, leads to the equations \eqref{hairy G00}--\eqref{hairy Bianchi}, with the various functions $A_{i,j}^{(\pm)}$, $B_{i,j}^{(\pm)}$, $C_{i,j}^{(\pm)}$, $D_{i,j}^{(\pm)}$ and $E_{i,j}^{(\pm)}$ defined as:
\begin{widetext}
    \begin{align}
        A_{i,0}^{(+)} &= \frac{1}{M_i^2}\sum_j\left(\beta_0^{(i,j)}U_i^2 + 2\beta_1^{(i,j)}U_i U_j + \beta_2^{(i,j)}U_j^2\right)
        \\
        A_{i,0}^{(-)} &= \frac{1}{M_i^2}\sum_k\left(\beta_4^{(k,i)}U_i^2 + 2\beta_3^{(k,i)}U_i U_k + \beta_2^{(k,i)}U_k^2\right)
        \\
        A_{i,j}^{(+)}=\frac{Y_i^2}{Y_j^2}A_{j,i}^{(-)} &= \frac{1}{M_i^2}\frac{Y_i}{Y_j}\left(\beta_1^{(i,j)}U_i^2 + 2\beta_2^{(i,j)}U_i U_j + \beta_3^{(i,j)}U_j^2\right)
        \\
        B_{i,0}^{(+)} &= \frac{1}{M_i^2}\sum_j\left[\left(\beta_0^{(i,j)}p_i+\beta_1^{(i,j)}p_j\right)U_i^2 + 2\left(\beta_1^{(i,j)}p_i+\beta_2^{(i,j)}p_j\right)U_i U_j + \left(\beta_2^{(i,j)}p_i+\beta_3^{(i,j)}p_j\right)U_j^2\right]
        \\
        B_{i,0}^{(-)} &= \frac{1}{M_i^2}\sum_k\left[\left(\beta_4^{(k,i)}p_i+\beta_3^{(k,i)}p_k\right)U_i^2 + 2\left(\beta_3^{(k,i)}p_i+\beta_2^{(k,i)}p_k\right)U_i U_k + \left(\beta_2^{(k,i)}p_i+\beta_1^{(k,i)}p_k\right)U_k^2\right]
        \\
        C_{i,0}^{(+)} &= \frac{1}{M_i^2}\sum_j \left[\left(\beta_0^{(i,j)}U_i + \beta_1^{(i,j)}U_j\right)p_i + \left(\beta_1^{(i,j)}U_i + \beta_2^{(i,j)}U_j\right)p_j\right]
        \\
        C_{i,0}^{(-)} &= \frac{1}{M_i^2}\sum_k \left[\left(\beta_4^{(k,i)}U_i + \beta_3^{(k,i)}U_k\right)p_i + \left(\beta_3^{(k,i)}U_i + \beta_2^{(k,i)}U_k\right)p_k\right]
        \\
        C_{i,j}^{(+)}=\frac{Y_i^2}{Y_j^2}C_{j,i}^{(-)} &= \frac{1}{M_i^2}\frac{Y_i}{Y_j}\left[\left(\beta_1^{(i,j)}U_i + \beta_2^{(i,j)}U_j\right)p_i + \left(\beta_2^{(i,j)}U_i + \beta_3^{(i,j)}U_j\right)p_j\right]
        \\
        D_{i,j}^{(+)}=D_{j,i}^{(-)} &= \beta_1^{(i,j)}U_i^2 + 2\beta_2^{(i,j)}U_i U_j + \beta_3^{(i,j)}U_j^2
        \\
        E_{i,j}^{(+)}=E_{j,i}^{(-)} &= 2\left[\left(\beta_1^{(i,j)}U_i + \beta_2^{(i,j)}U_j\right)p_i + \left(\beta_2^{(i,j)}U_i + \beta_3^{(i,j)}U_j\right)p_j\right]
    \end{align}
\end{widetext}

\clearpage
\bibliography{bibliography.bib}
\bibliographystyle{JHEP}
\end{document}